\documentclass[prb,aps,amssymb,showpacs,twocolumn,amsmath,floatfix]{revtex4}
\usepackage{tabularx}
\usepackage{bm}
\usepackage{euscript}
\usepackage{epsfig,psfrag,subfigure}
\usepackage{graphicx}
\usepackage{color}
\usepackage{amsfonts}
\usepackage{exscale}
\usepackage{amsbsy}
\usepackage{hyperref}
\input{epsf}

\def\be{\begin{equation}}
\def\ee{\end{equation}}
\def\ba{\begin{eqnarray}}
\def\ea{\end{eqnarray}}

\begin{document}

\title{Superconductivity in the repulsive Hubbard model: an asymptotically exact weak-coupling solution}
\author{S. Raghu$^1$, S. A. Kivelson$^1$ and D. J. Scalapino$^{1,2}$}
\affiliation{$^1$Department of Physics, Stanford University, Stanford, CA 94305-4045}
\affiliation{$^2$Department of Physics, UCSB, Santa Barbara, CA 93106-9530}
\date{\today}

\begin{abstract}
We study the phase diagram of the Hubbard model in the limit where U,
the onsite repulsive interaction, is much smaller than the bandwidth.  We present an
asymptotically exact expression for T$_c$, the superconducting transition temperature,
in terms of the correlation functions of the non-interacting system which is
valid for arbitrary densities so long as the interactions are sufficiently small.   Our strategy for
computing T$_c$ involves first integrating out all degrees of freedom having energy higher
than an unphysical initial cutoff $\Omega_0$.  Then, the renormalization group (RG) flows of the
resulting effective action are computed and T$_c$ is obtained by determining the scale
below which the RG flows in the Cooper channel    
diverge.  We prove that T$_c$ is independent of $\Omega_0$.  Using this method,
we find a variety of unconventional superconducting ground states in two and three dimensional lattice systems and present explicit results for T$_c$ and pairing symmetries as a function of  the electron
concentration.  
\end{abstract}


\maketitle

\section{Introduction}
The Hubbard model 
 is widely studied as the paradigmatic model of strongly correlated electrons\cite{Anderson1987,Hubbardbook}.  However, in more than one dimension (1D) there is controversy  concerning even the basics of the phase diagram of the model.  Most theoretical work on the model has focused on intermediate to strong interactions, $U \sim W$, since this is the physically relevant range of parameters for any of the intended applications of the model to real solid state systems.  (Here, $U$ is the repulsion between two electrons on the same site, and $W$ is the bandwidth in the limit $U=0$.)  However, for such strong interactions the only well controlled solutions are numerical 
 and the application of determinental quantum Monte Carlo methods\cite{Blankenbecler1981} and the Density-Matrix-Renormalization-Group\cite{White1992} have been limited by the fermion sign\cite{Troyer2005} and two-dimensional entanglement problems\cite{Hastings2007} respectively.

Here, we study  the  limit of weak interactions, $U/t \to 0$, where we compute the phase diagram and obtain expressions for the critical temperatures which, assuming the validity of certain assumptions discussed below,  are asymptotically exact.  To be explicit, we consider the Hubbard model  
\begin{eqnarray}
H=&& H_0 + U \sum_i c^\dagger_{i\uparrow}c^\dagger_{i,\downarrow}c_{i,\downarrow}c_{i,\uparrow} \\
\label{Hubbard}
H_0=&&-t \sum_{<i,j>,\sigma}[c^\dagger_{i,\sigma}c_{j,\sigma}+ h.c.] - t' \sum_{(i,j),\sigma}[c^\dagger_{i,\sigma}c_{j,\sigma}  + h.c.]
\nonumber
\end{eqnarray}
for a variety of lattice systems in two and three dimensions.  Here, 
$c^\dagger_{i,\sigma}$ creates an electron with spin polarization $\sigma$ on lattice site $i$, and  $<i,j>$ and $(i,j)$ signify, respectively, pairs of nearest-neighbor and next-nearest-neighbor sites.  

Since the Cooper instability is the only generic instability of a Fermi liquid, except for certain fine tuned values of $t'/t$ and the electron density $n$, the only ordered states that can be stabilized by weak interactions are superconducting states.  For repulsive interactions, $W \gg U >0$, the superconducting transition temperature has an asymptotic expansion 
\ba
&&T_c \sim W\ \exp\left\{ -\alpha_2 (t/U)^2 -\alpha_1 (t/U) -\alpha_0 \right\}\nonumber \\
&&\ \ \ \ \ \ \ \ \ \times \Big [ 1 + {\cal O}(U/t) \Big] \nonumber\\
&&\sim W \exp\left\{-1/\left[\rho V_{\rm eff} \right] \right\} \left[ 1 + {\cal O}(U/t) \right] 
\label{Tc}
\ea
where $\alpha_n$ are dimensionless functions of $t'/t$, $n$ and $\rho$ is the density of states at the Fermi energy.  The principal result we report here is to give an explicit prescription for computing $\alpha_2$ and  $\alpha_1$ as a function of the electron density, $n$, and the ``band structure".  
On the basis of the present analysis, we conclude that the resulting phase diagram is asymptotically exact in the sense that 
\begin{equation}
\lim_{U\to 0} \left\{(U/t)^2 \ln[W/T_c]\right\} = \alpha_2.
\end{equation}
We will also explain why we are unable to give a prescription for computing $\alpha_0$.  In the process of computing $\alpha_2$, one determines the symmetry of the superconducting ground state (e.g. s-wave, p-wave, d-wave, etc.) and the form of the pair wavefunction.  


There are, of course, special situations in which a variety of different non-superconducting ordered phases occur.  While these situations are potentially significant in what they imply about the behavior of the system at intermediate $U$, in the small $U$ limit they always involve a large degree of fine tuning of parameters.   
The canonical example is the case of a square lattice, in which the model with $t'=0$ has a non-generic particle-hole symmetry which leads to perfect nesting of the Fermi surface
when the mean electron density per site is $n=1$, where
\begin{equation}
n\equiv N^{-1} \sum_{j\sigma}<  c^\dagger_{j,\sigma}
c_{j,\sigma}> ,
\end{equation}
 These special situations are thus, in some sense, not really a part of the weak coupling problem, but rather a piece of the strong correlation problem that persists to weak coupling.  
  If one's principal interest\cite{Luther1994,Fjaersestad1999,Balents1996,Lin1998,Dimov2008} is in extrapolating well controlled weak-coupling calculations to the range of strong interactions, this degree of fine tuning of the bandstructure is a small price to pay to gain access to phases which have broad ranges of stability for intermediate to large $U$.  However, if we focus on the small $U$ limit in its own right, then for any fixed, non-zero value of $t'/t$ no antiferromagnetic insulating phase occurs. 

 
 An interesting interplay between various possible ordered phases can also occur when  the Fermi energy is coincident with a van-Hove singularity\cite{Dzyaloshinskii1987,Schulz1987,Lederer1987,Zanchi1996, Zanchi2000}.  While these singularities occur for generic values of $t'/t$, the density must be fine tuned in order for the singularity to lie sufficiently close to the Fermi energy to matter.  
 The study of the behavior of the system in weak coupling tuned near a vanHove singularity has been explored by several authors\cite{Halboth2000, Honerkamp2002,Neumayr2003,Khavkine2004,Yamase2005,Kee2005}, again as a route to understanding strong-coupling phases and the interplay between phases in a regime of parameters in which perturbative renormalization group (RG) methods can be applied.  However, again, as we are focussing on  the physics of a system with small $U/t$, we do not treat the interplay with non-superconducting orders.
 
 The original idea of obtaining superconductivity from  repulsive interactions dates back to 
the pioneering work of Kohn and Luttinger\cite{Kohn1965} who derived an effective attractive 
interaction from the Friedel oscillations of a 3D electron gas.  
RPA calculations for a repulsive Hubbard model found that near a 
spin-density-wave instability there was an effective interaction 
which favored d-wave superconductivity\cite{Scalapino1986}.  
The treatment of this problem from the 
standpoint of the renormalization group was presented by Zanchi and Schulz\cite{Zanchi1996, Zanchi2000} and others\cite{Salmhofer1998,Binz2003,Halboth2000,Honerkamp2001}.  Furthermore, the problem of competing instabilities of electronic systems has extensively been studied via the numerical functional renormalization group (FRG) methods\cite{Honerkamp2001a,Honerkamp2002,Metzner2006,Zhai2009}.    While  these works have made significant 
progress in our understanding of superconduvtivity from repulsive interactions, we present here an 
 asymptotic analysis of the problem and show explicitly the way the final expressions for the 
superconducting transition temperature are  independent of the initial choice of cutoff.  
Our analysis is based on the renormalization group framework established by Shankar\cite{Shankar1994} and Polchinsky\cite{Polchinski1992}.
In section \ref{relation}, we present a more complete discussion of the relationship between the work presented here and previous analyses of this problem.

 This paper is organized as follows.  In section II, we present the results of our perturbative 
 RG treatment.  Expressions for the quantity $\alpha_2$ are presented in the spin singlet and triplet 
 pairing channels.  These results are then applied to a variety of systems in section III.  
 Both lattice and continuum systems in $d=2$ and $d=3$ are considered.  In section IV, we present the 
overall strategy of our RG calculations.  In section V, we  describe the technical aspects of the perturbative renormalization of the effective 
 interactions in the Cooper channel, and discuss the one-loop RG flows of the effective pairing vertex in section VI.  
 In section \ref{relation} we discuss the relation of the present work to previous closely related approaches, and we conclude, in section \ref{discussion} with a few remarks about future directions.  There is an appendix with technical details.
 
  To simplify our notation, we henceforth adopt units in which  $\hbar=1$, and the volume of the unit cell $\nu=1$. 
 
 \section{Results}
 
 Before discussing the derivation, we articulate the final results of our analysis, {\it i.e.} we give an explicit method to compute $\alpha_2$ exactly.  (It is more complicated to compute $\alpha_1$, and since it is subdominant; we defer that part of the discussion to the more technical portion of the paper.)  The results are expressed in terms of $\chi(\vec k)$, the static susceptibility of the non-interacting system,
 \begin{equation}
\chi(\vec k) = - \int \frac{d^d q} {(2\pi)^d} \frac {[ f(\epsilon_{\vec k+\vec q}) - f(\epsilon_{\vec q})] }
{[\epsilon_{\vec k+\vec q} - \epsilon_{\vec q} ]}
\end{equation}
evaluated in the limit $T\to 0$, and 
\begin{equation}
\rho = \int \frac{d^d q} {(2\pi)^d}\delta(\epsilon_{\vec q}) = \lim_{|\vec k| \to 0} \chi(\vec k) ,
\end{equation}
 the density of states at the Fermi energy E$_f$.   Here $\epsilon_{\vec{q}}$
is the band-dispersion measured relative to the Fermi energy, $E_F$, and $d=2$ or 3 denotes the number of spatial dimensions.   
The integrals run over the appropriate first Brillouin.  In general, $\chi$ must be computed numerically, but this is a straightforward computation.

For simplicity, we shall restrict our attention in this paper to systems which possess inversion symmetry and no spin-orbit coupling.  This 
enables us to classify the possible superconducting states as having either even or odd parity, the former class consisting of spin singlet  and the latter of spin triplet states.  
Among the even parity spin-singlet superconducting states, the s-wave states are those which transform trivially under a point group operation of the crystal, whereas the ``d-wave" and higher angular 
momentum channel gap functions transform according to a non-trivial representation of the point 
group.  The spin triplet states include p-wave and f-wave gap functions, all of which transform 
according to a non-trivial representation of the point group.  Depending on the crystalline point group,  
the superconducting gap functions may transform as either a non-degenerate or an n-fold degenerate 
irreducible representation of the point group operations.  We will present expressions for $\alpha_2$ for 
each of these states; for any given bandstructure and electron concentration, the physical low 
temperature phase is that one which produces the smallest  value of $\alpha_2$.  


In order to compute $\alpha_2$, we must solve, for each symmetry class, the eigenvalue problem displayed below.  In the spin-singlet channel, the eigenvalue problem corresponds to the integral 
equation
\begin{eqnarray}
&& \int \frac {d\hat q}{S_F}\ \bar g^s_{\hat k,\hat q}\ \psi^{(n)}_{s, \hat q} = \lambda_n  \psi^{(n)}_{s, \hat k} \nonumber \\
&&  \bar g^s_{\hat k,\hat q} = \rho U^2 \sqrt{\frac {\bar v_F}{v_F(\hat k)}} \left[ \chi(\hat k + \hat q) + c_1 \right] \sqrt{\frac {\bar v_F}{v_F(\hat q)}} 
\label{singlet}
\end{eqnarray}
where $c_1$ is 
of order $t/U$, $\hat k$  designates a vector on the unperturbed Fermi surface, $S_F\equiv\int d\hat p$ is the ``area'' of the Fermi surface,
$v_F(\hat p)$ is the magnitude of the Fermi velocity at position $\hat p$, and the norm of the Fermi velocity is defined according to
\begin{equation}
\frac 1 {\bar v_F} \equiv \int \frac {d\hat p}{S_F}\left(\frac {1}{v_F(\hat p)}\right).
\end{equation}
The only effect of $c_1$ is to penalize the trivial 
s-wave state due to the bare onsite repulsive interaction.  Eigenfunctions with higher angular 
momentum, such as the d-wave states,  
and appropriate ``extended s-wave'' states are unaffected by it\cite{Pitaevskii1960,Brueckner1960,Emery1960}.   
In the spin-triplet channel, the eigenstates obey
\begin{eqnarray}
&& \int \frac {d\hat q}{S_F}\ \bar g^t_{\hat k,\hat q}\ \psi^{(n)}_{t, \hat q} = \lambda_n  \psi^{(n)}_{t, \hat k} \nonumber \\
 && \bar g^t_{\hat k,\hat q}= -\rho U^2 \sqrt{\frac {\bar v_F}{v_F(\hat k)}} \ \chi(\hat k - \hat q)\ \sqrt{\frac {\bar v_F}{v_F(\hat q)}}
\label{triplet}
\end{eqnarray}

Assuming that there  is at least one negative eigenvalue present in either Eq. \ref{singlet} or \ref{triplet}, the quantity $\alpha_2$ is obtained by the relation  
\begin{equation}
\alpha_2 = |\lambda_0|^{-1}
\left(U/t\right)^2, 
\end{equation}
where $\lambda_0$ is the most negative eigenvalue.    The zero-temperature 
gap function is proportional to $\psi$:
\begin{equation}
\Delta_{s(t)}(\hat{k})\sim T_c 
\sqrt{\frac{v_f(\hat{k})}{\bar{v}_f}} \psi_{s(t)}(\hat{k}).
\end{equation}
The computation of $\alpha_1$ involves an analysis of less singular scattering processes in 
perturbation theory and will be addressed in section VI.  

\section{Application to various systems}
The expressions in Eqs. \ref{singlet} and \ref{triplet} hold for the Hubbard model and are independent of 
the microscopic details of the electronic structure.  
In this section we determine the superconducting ground states for  a 
variety of inversion and spin-rotationally symmetric systems in $d=2$ and $d=3$.  

There are two distinct effects of the band-structure which affect the asymptotic behavior of $T_c$:  Firstly, the existence of a superconducting instability from repulsive interactions at all derives from the $\vec k$-space structure of the effective interactions, and these depend strongly on $n$ and the details of the band structure.   Secondly, the dimensionless density of states at the Fermi energy, $\rho t$, varies  (in 2d especially) with distance from a van-Hove point.  To distinguish these two effects, in the figures we re-express the leading-order asymptotics as
\be
T_c \sim \exp\left\{-[\rho V_{\rm eff}]^{-1}\right\} 
\ee
where
\be
V_{\rm eff} (n) = \vert \lambda \vert /\rho = \left( U/t \right)^2[\alpha_2 \rho  ]^{-1}.
\ee
\subsection{Rotationally invariant systems in $d=2$ and $3$}
We begin by considering electrons in the continuum limit 
with quadratic dispersion (which is achieved in a lattice system 
in the limit $n \ll 1$): 
\begin{equation}
\epsilon_{\vec k} = \frac{k^2}{2 m}.
\end{equation}
In 3 dimensions, 
\begin{eqnarray}
\chi(q) &=& \rho \left[ k_f + \frac{\left(2 k_f \right)^2 - q^2}{4 q} \log { \left | \frac{q + 2 k_f}{q-2k_f} \right | } \right] \nonumber \\
& \simeq & \rho \left( 1 - \frac{1}{2} \left( \frac{q}{2 k_f} \right)^2 + \cdots \right)
\end{eqnarray}
where $\rho = m k_f/2 \pi^2$.  
For such a rotationally-invariant system, each eigenfunction can be classified according to its angular momentum $\ell$, and the solutions are $\left(2 \ell + 1 \right)$-fold degenerate.   The expansion 
of $\chi(q)$ in powers of $q/2k_f$ is justified when seeking the lowest angular momentum pairing 
solutions and it shall suffice to compare the s-wave and p-wave solutions which require only the 
leading order expansion about $q = 0$.  
The non-degenerate s-wave eigenfunction has a uniform gap everywhere on the Fermi surface.
From Eq. \ref{singlet}, it can easily be seen that  since the susceptibility is a positive-definite quantity, a negative eigenvalue for an s-wave gap function does not exist.   However,  the 3-fold degenerate p-wave channel with {\it e.g.} $\psi_t(\theta) \propto \cos{\theta}$, does have a negative eigenvalue, as can be seen by applying Eq. \ref{triplet}:  
\begin{equation}
-\rho \int \frac{ d \Omega'}{4 \pi} \left( 1 - \frac{1}{2} \left( \frac{\vert \hat{k}-\hat{k}' \vert}{2 k_f} \right)^2 \right) \psi_t(\Omega')= \lambda \psi_t (\Omega) \nonumber \\
\end{equation}
The solution is $\lambda = - \rho/16 < 0$. 
Thus, for a system with a quadratic dispersion and a spherical Fermi surface, the ground state is a p-wave superconductor  to order $U^2$.    Physically, this can be understood from the fact that the dominant spin fluctuations are ferromagnetic.   Indeed, such considerations have been 
applied extensively to superfluid Helium-3 \cite{Anderson1975, Leggett1975}.

For a two-dimensional rotationally invariant system with quadratic dispersion, 
\begin{equation}
\chi(q) =  \frac{m}{2 \pi} \left[  1- \frac{ {\rm Re} \sqrt{ q^2 - (2 k_f)^2} }{q}  \right] .
\end{equation}
Note that the susceptibility is a constant for $q < 2k_f$.  However, for a system with a circular Fermi surface, the furthest apart any two points on the Fermi surface can be is $2k_f$.  
Therefore, the pairing strengths correspond to the eigenvalues of a constant matrix.  
One sees immediately from this that the s-wave solution has a positive eigenvalue $\lambda = m/2 \pi + c_1$, 
and all higher angular momentum gap functions have zero eigenvalues.  Therefore, for such a 
rotationally invariant system in two dimensions, superconductivity does not occur in any channel 
to order $U^2$.  
When higher order terms in perturbation theory are taken into account, 
it has been found that the leading instability occurs in the p-wave channel\cite{Chubukov1993}.  

\subsection{Lattice systems: d=2}
Various authors\cite{Baranov1992, Chubukov1992, Hlubina1999, Fukagawa2002} have studied the Hubbard model on the 2D square lattice in the weak 
coupling limit.  Near half-filling, nesting effects lead to $d_{x^2-y^2}$ pairing.  In weak coupling 
at low doping, higher order non-quadratic terms in the quasiparticle dispersion\cite{Baranov1992} lead to $d_{xy}$ 
pairing for $t'=0$ and for $t' < -t/4$, the p-wave state is favored\cite{Chubukov1992}.  In higher order perturbation 
theory\cite{Fukagawa2002}, $U^3$ vertex corrections have been shown to enhance the p-wave over the $d_{xy}$ state.  
Here, we are interested in the behavior of the model as $U/t$ approaches zero while keeping 
the density of electrons fixed, and we have studied  the eigenvalue problems given in Eqs. \ref{singlet} and  \ref{triplet} 
must be studied numerically.  
A finite number of points on the Fermi surface form the basis for the matrices  $g^{s,t}$.  These in turn 
are diagonalized and the pairing eigenvalues are determined for each electron concentration.  
For each of the systems considered below, we have found that a discretization of the Fermi surface with $500-800$ points is more than sufficient to produce accurate results.  Both 
spin and the lattice 
point group symmetries are used to classify the resulting eigenstates.  For a tetragonal crystal (i.e. the point group D$_{4h}$) without 
spin-orbit coupling, there are 4 non-degenerate spin singlet states, and a two-fold degenerate triplet 
state:
\begin{eqnarray}
A_{1g}: \psi &\sim& 1, {\rm \ or \ } \left(x^2+y^2 \right) \nonumber \\
A_{2g}: \psi &\sim& (x^2-y^2)xy \nonumber \\
B_{1g}: \psi &\sim& (x^2-y^2) \nonumber \\
B_{2g}: \psi &\sim& xy \nonumber \\
E_u : \psi & \sim& \left\{ x,y \right\}
\end{eqnarray}
where the left hand side labels the irreducible representation and the right hand side lists the 
basis functions (with the association $x \rightarrow \sin{k_x}, x^2 \rightarrow \cos{k_x}$, etc).  

\begin{figure}
\includegraphics[width=3.5in]{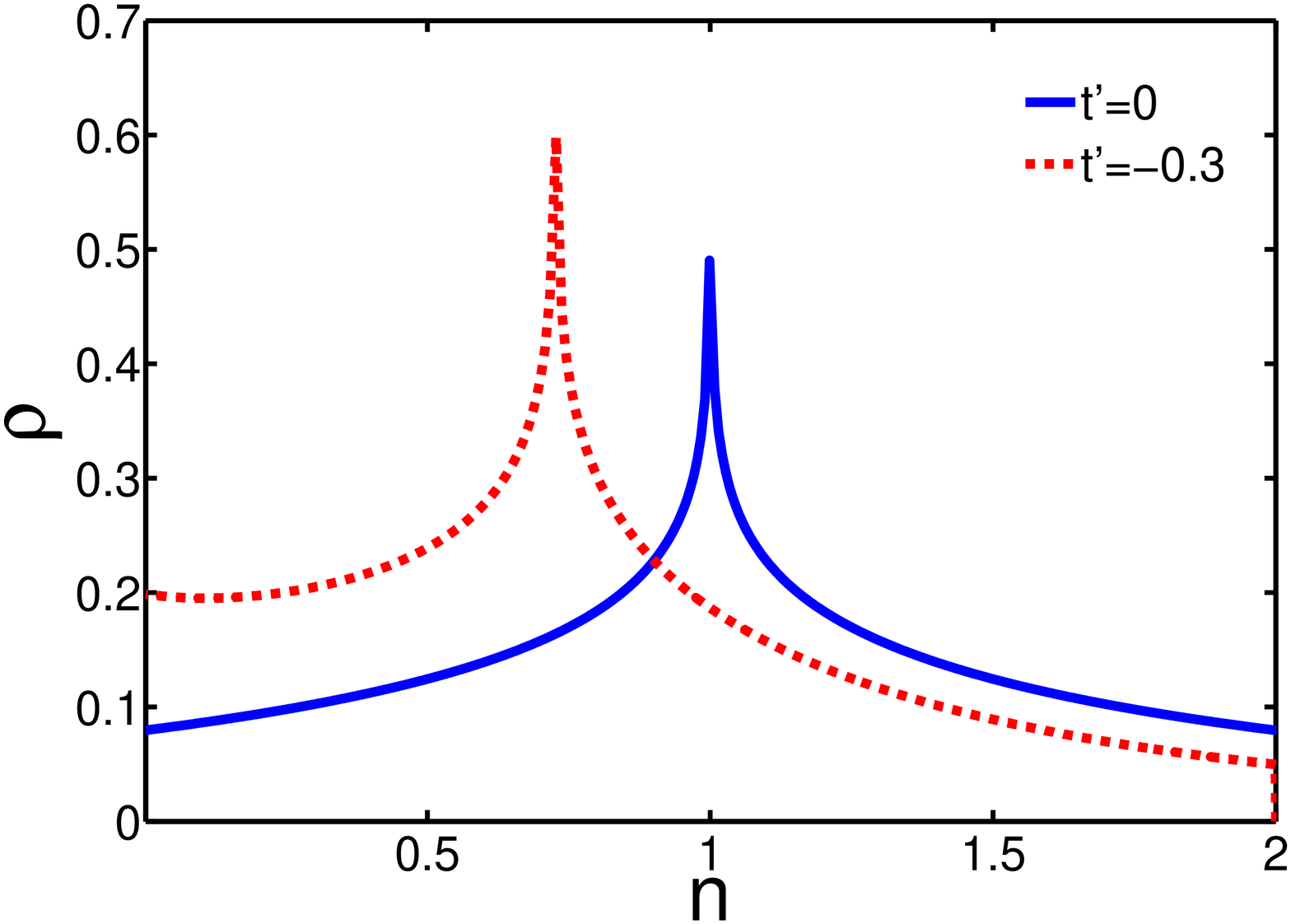}
\caption{Density of states as a function of electron concentration on the square lattice 
for $t'=0$ (solid line) and $t'=-0.3$ (dashed line).  
}
\label{dossquare}
\end{figure}

\begin{figure}
\includegraphics[width=3.5in]{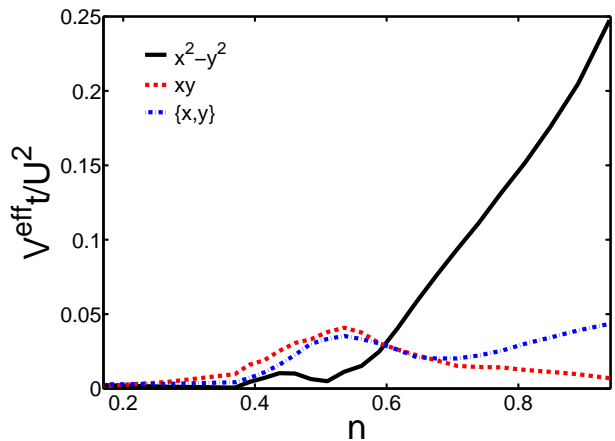}
\caption{Pairing strengths for the 2D Hubbard model at $t'=0$ as a function of 
electron concentration.    
}
\label{tprime0}
\end{figure}

\begin{figure}
\includegraphics[width=3.0in]{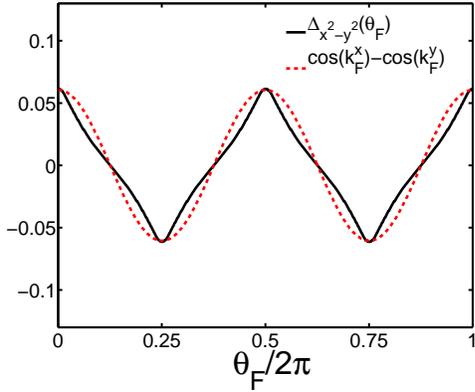}
\caption{The gap function $\Delta$ for the $d_{x^2-y^2}$ state which occurs at $t'=0$ and 
$n=1.1$. The pair field is plotted as a function of $\theta_F$, the angular degree of freedom on the Fermi surface relative to the $x$ axis.  The pair field obtained in the weak coupling analysis deviates 
significantly from a simple $\cos(2 \theta_F)$ form, depicted here by the dotted line.  }
\label{wf}
\end{figure}

Figure \ref{dossquare} shows $\rho$ as a function of electron concentration $n$ for a 
tight-binding model on a square lattice.  
Figure \ref{tprime0} shows the pairing strengths for the 2D Hubbard model on a square lattice with $t' = 0$ as a function of 
$n$.  
(Particle-hole symmetry assures that the phase diagram is invariant under $n\to 2-n$.)  We see clearly that near half-filling ($n=1$), the dominant form of superconductivity has $d_{x^2-y^2}$ symmetry.  At about 
$n=0.6$, we see that the favored configuration changes to $d_{xy}$ pairing.  
Thus, there are two distinct superconducting ground states that occur on a square lattice with a near-neighbor hopping as a function of concentration for the particle-hole invariant system at $t'=0$: there is a $d_{x^2-y^2}$ ground state for $1 > n > 0.6$,
and a $d_{xy}$ ground state for 
$n <  0.6$.  
The results shown in Fig. \ref{tprime0} are in  qualitative agreement with the 
spin fluctuation exchange studies of the Hubbard model of Scalapino {\it et al}\cite{Scalapino1986}.  
Figure \ref{wf} shows the gap function for the d-wave state at 
$n=1.1$.  Notice that it has a shape that differs substantially from the simple $\left( \cos{k_x} - \cos{k_y} 
\right)$ form.   

 \begin{figure}
\includegraphics[width=3.5in]{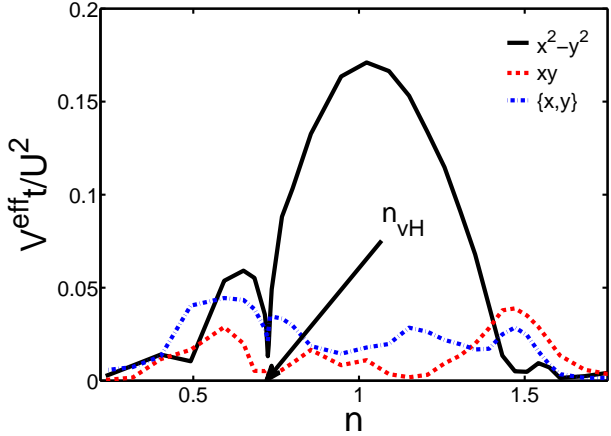}
\caption{Pairing strengths for the 2D Hubbard model at $t'=-0.3t$ as a function of electron density.  
The critical density $n_c$ at which the van Hove singularity occurs at the Fermi level is shown.   }
\label{tprime.3}
\end{figure}

For $t' \ne 0$, on the square lattice, the particle-hole symmetry is destroyed and the van Hove singularities occur away from half-filling.  Figure \ref{tprime.3} shows the pairing strengths on the square lattice at $t' = -0.3$ as a function of doping.   Again, the dominant configuration which occurs near the 
half-filled system is $d_{x^2-y^2}$ pairing.  The van Hove singularity occurs in this system at 
$n=n_{vh}\approx 0.72$.  In a narrow window of densities near $n_{vH}$, the d-wave order is suppressed and the p-wave order dominates.  
Upon further decreasing the electron concentration away from the van Hove singularity, the $d_{x^2-y^2}$ ground state again gives way to a p-wave superconducting state at 
$n\approx 0.5$.  Qualitatively similar 
results have been found for the square lattice $t,t'$ Hubbard model using functional renormalization 
group analysis for fillings close to the van Hove singularities \cite{Honerkamp2004}.  

\begin{figure}
\includegraphics[width=3.5in]{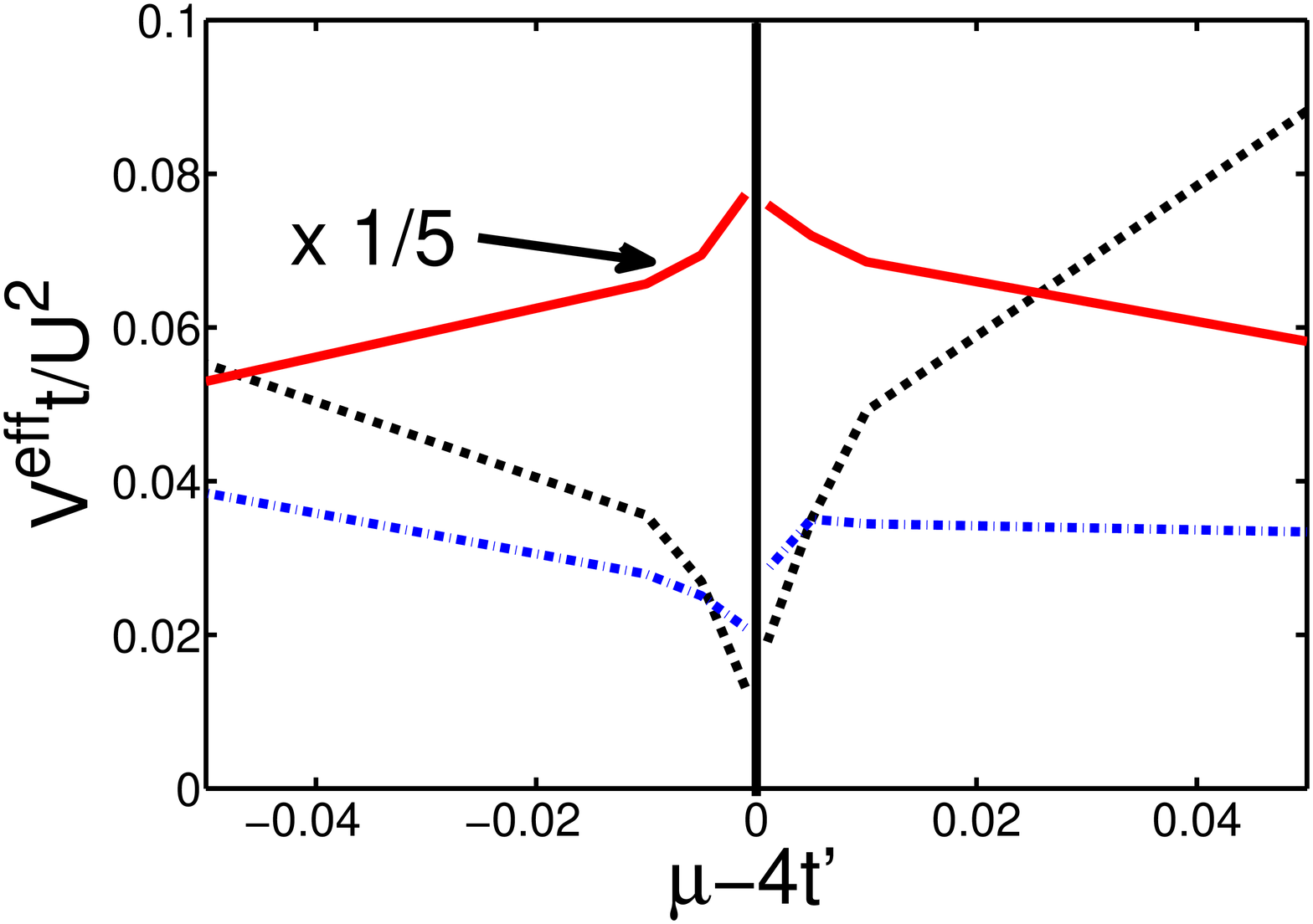}
\caption{The effective interaction for $t'=-0.3$ (dashed and dashed-dotted lines) and for $t'=-0.05$ (solid line) as a function of $\mu-4t'$.  For $t'=-0.3$, the dashed line represents the pairing strength of the $d_{x^2-y^2}$ state whereas the dashed-dotted line corresponds to the pairing strength of the p-wave state.  While both are depressed as the chemical potential crosses the van Hove  singularity, the p-wave state 
obtains a larger pairing strength parametrically close to the van Hove point.  For $t'=-0.05$, the 
nesting of the Fermi surface remains nearly perfect and acts to enhance the pairing strength of the $d_{x^2-y^2}$ state.  The p-wave pairing strength for $t'=-0.05$ is not shown here, since it is much lower 
than the d-wave strength.    }
\label{vh}
\end{figure} 
It is worth examining the singular behavior near $n_{vH}$ (i.e. near the point $\mu = 4t'$) in more 
detail.  
For the square lattice with $t' \ne 0$, the susceptibility at $\vec{q} = \left(0,0 \right)$ diverges 
logarithmically as $\mu \rightarrow 4 t'$:
\begin{equation}
\chi(0) \sim \frac{1}{2 \pi^2 t} \ln\left|\frac{t}{\mu-4t'}\right|
\end{equation}
For a large 
momentum transfer, $\vec{Q} = \left(  \pi, \pi \right)$, the susceptibility varies as 
\begin{equation}
\chi(\vec{Q}) \sim \frac{1}{2 \pi^2 t} \ln\left|\frac{t}{\mu-4t'}\right| \ln \left| \frac{t}{2 t'} \right|
\end{equation}
for $\vert \mu - 4 t' \vert << t'$.  Thus, a finite $t'$ acts to suppress the nesting of the Fermi surface (
for the perfectly nested Fermi surface at $t' = 0$, $\chi(\vec{Q})$ diverges much more strongly as  $\ln^2 \vert t/\mu \vert$).  Taking the dominant scattering processes  at momenta $\vec q = \left(0,0 \right)$ and $\vec Q = \left( \pi, \pi \right)$ into account, one finds that the effective d-wave pairing strength near 
the van Hove singularity is roughly 
\begin{equation}
V_{eff}  \simeq \frac{U^2}{2 \pi^2 t} \ln \left| \frac{t}{\mu-4t'} \right| \left( \ln \left| \frac{t}{2t'} \right| - 1 \right) + V_0
\end{equation}
where $V_0$ is a subdominant contribution which arises from the intermediate momentum transfers on the Fermi surface.  For $t' << t$, the d-wave pairing strength is enhanced as $\vert \mu - 4 t' \vert \rightarrow 0$.  However, for $t/2t' > e$ the d-wave pairing strength {\it decreases} as the van Hove 
singularity is approached.  This behavior is illustrated in Fig. \ref{vh}.  It reflects the fact that the d-wave 
pairing is driven by an interaction which is stronger at large momentum transfer.  An interaction which is 
greater at small momentum transfer suppresses the d-wave pairing.  

\begin{figure}
\includegraphics[width=3.0in]{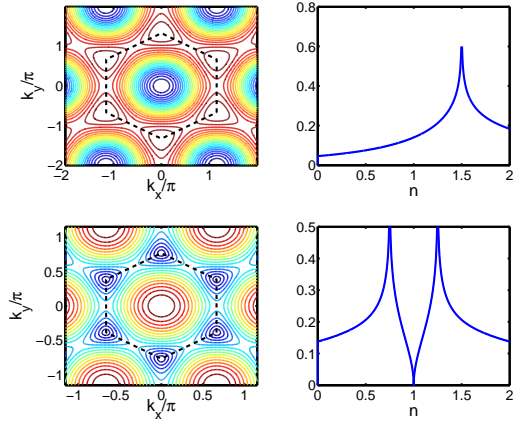}
\caption{(upper left): Energy contours of the triangular lattice nearest-neighbor 
tight-binding model.  The dashed black line marks the zone boundary.  Blue contours correspond to the band bottom and red contours 
occur near the top of the band.  (upper right) Density of states on the triangular lattice.  When 
the Fermi level is at the van Hove point, the volume of the Fermi surface is 3/4 of the zone.  (lower left) 
Energy contours of the honeycomb lattice nearest-neighbor tight-binding model.  Due to the particle-hole symmetry, only the $\epsilon >0$ contours are shown.  (lower right) Density of states of the honeycomb model.  The van Hove singularities occur at a filling of 3/8 and 5/8.  }
\label{trihc}
\end{figure}

\begin{figure}
\includegraphics[width=3.5in]{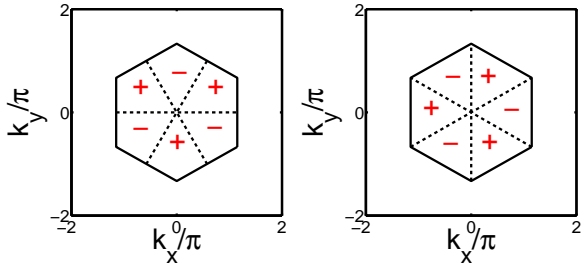}
\caption{ The two distinct f-wave irreducible representations on the triangular and honeycomb 
lattices.  (left) B$_{1u}$: $\psi \sim   k_y \left( k_y^2 -3k_x^2 \right)$.  (right)  B$_{2u}$: $\psi \sim   k_x \left( k_x^2 -3k_y^2 \right)$.   The B$_{1u}$ gap function is the 
dominant gap when there are two disconnected Fermi pockets in the system, each centered around 
the zone corners.  The B${1u}$ gap function has the opposite 
sign on each pocket.}
\label{fgaps}
\end{figure}

Next we consider the nearest-neighbor tight-binding model on the 2D 
triangular and honeycomb lattices.  Figure \ref{trihc} shows the basic electronic 
structure on these lattices.  Both lattice systems have the hexagonal point group (D$_{6h}$) symmetry
and therefore the same irreducible representations characterize their gap functions.  
In the singlet (even parity) channel, the following are the 
irreducible representations:
\begin{eqnarray}
{\rm Singlet \ channel} \nonumber \\
A_{1g}: & & \psi \sim k_x^2 + k_y^2 \nonumber \\
E_{2g}: & & \psi \sim \left\{ \left( k_x^2 -k_y^2 \right),  \left( 2k_x k_y \right) \right\} \nonumber \\
A_{2g}: & & \psi \sim k_x k_y (k_x^2-3k_y^2)(k_y^2-3k_x^2) \nonumber \\
\end{eqnarray}
Note that the d-wave function is a two-dimensional representation (this is in fact the only 
two dimensional representation in the singlet channel for this system, so long as we restrict our superconductivity to be 
only in the basal plane).  For the triplet channel, we have
\begin{eqnarray}
{\rm Triplet \ channel} \nonumber \\
E_{1u}: & & \psi \sim \left\{ k_x , k_y \right\} \nonumber \\
B_{1u}: & & \psi \sim k_y \left( k_y^2 -3k_x^2 \right) \nonumber \\
B_{2u}: & & \psi \sim k_x \left( k_x^2 -3k_y^2 \right) \nonumber \\
\end{eqnarray}
The p-wave gaps form a  two-dimensional irreducible representation.   There are two distinct one-dimensional f-wave 
gap functions that belong to the B$_{1u}$, B$_{2u}$ representations and are shown in  Fig. \ref{fgaps}.  

\begin{figure}
\includegraphics[width=3.5in]{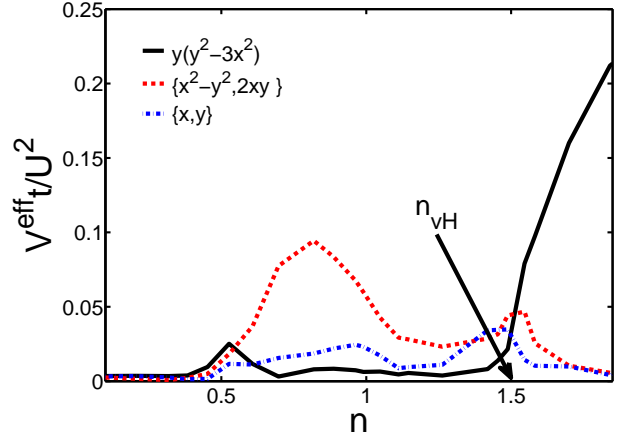}
\caption{ The dominant pairing strengths on the triangular lattice as a function of electron concentration.  The two-component d-wave order parameter 
occurs in the hole-doped (i.e. $x < 0$) system.  But for electron-doping ($x>0$) beyond the van-Hove filling, the f-wave(2) gap develops the largest critical temperature.   }
\label{triangulartc}
\end{figure} 

\begin{figure}
\includegraphics[width=3.5in]{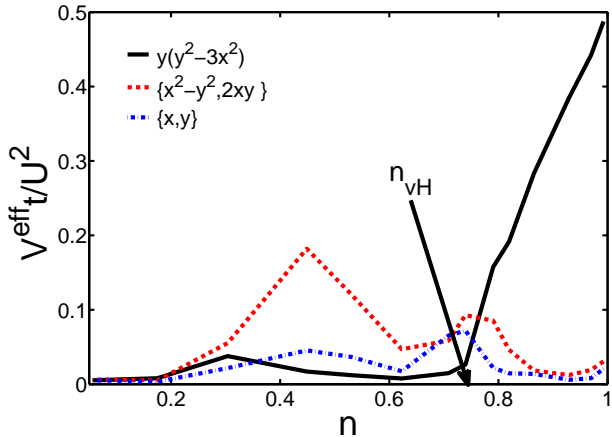}
\caption{Pairing strengths for the honeycomb lattice.  The trends on the honeycomb lattice are some what reverse in comparison to the triangular lattice.  However this can be understood from looking at the Fermi energy contours.  For a given sign of $t$, the energy contours of  honeycomb lattice for large electron doping (i.e. top of the band) is similar to those of  the triangular lattice at large hole doping (i.e. bottom of the band) and {\it vice-versa}. }
\label{honeycombtc}
\end{figure} 

Figs \ref{triangulartc} and \ref{honeycombtc} show the pairing strengths for the triangular 
and honeycomb lattice respectively.  For both systems it is seen that the dominant pairing instabilities are either the two-component d-wave representation, or the non-degenerate f-wave representation.  

For the triangular lattice at modest electron concentration, or the honeycomb lattice far away from half-filling, the Fermi surface is hexagonal in shape and is simply connected.  In this regime, the dominant pairing configuration is d-wave pairing that consists of a $d_{x^2-y^2}$ component degenerate with a $d_{xy}$ component.\cite{Doniach2007}  In our analysis, the magnitude of each of the two d-wave gap functions and their relative  phase  cannot be determined, since they are determined by non-linear effects captured, for instance in the Landau Ginzburg theory.  However, it is reasonable to expect that in order to gain condensation energy, the system will spontaneously break time-reversal symmetry and form a $d+id$ superconductor.    

For the triangular lattice near the top of the band, and for the honeycomb lattice close to half-filling, the 
Fermi surfaces form disjoint pockets.  In this concentration regime, we have found for both systems that 
the triplet f-wave gap function is the ground state.  The f-wave state which is favored has its lines of 
nodes along the lines connecting the zone center to the midpoints of the zone edges (see Fig. \ref{fgaps}).  However, since these lines of nodes never cross the Fermi surfaces which are centered on the 
corners of the Brillouin zone, the f-wave pairing produces a fully gapped state on the Fermi surface.  The gap changes sign between the two distinct Fermi surfaces.  It has been argued in the past based on the 
spin-fluctuation exchange mechanism that the strong magnetic excitations associated with such 
disjoint, reasonably well-nested Fermi pockets results in an effective pairing interaction 
which is repulsive between the two pockets.  Consequently, a solution which encodes a sign change 
of the gap among the two Fermi surfaces will naturally be favored \cite{Kuroki2001}. 
The transition between the f-wave and d-wave pairing states occurs as the electron density is 
varied across the van Hove filling.  We note, finally, that the f-wave solution has a substantially 
higher transition temperature than the other superconducting gap functions found in these lattice 
systems.

\subsection{Lattice systems: d=3}
Next we consider 3 dimensional lattice systems.  In particular, we shall consider the superconducting 
instabilities of the Hubbard model on the simple cubic (SC), body-centered cubic (BCC), face-centered 
cubic (FCC) and diamond lattices.  We shall restrict our analysis to nearest neighbor tight binding dispersion in each of these cases.  All of these lattices have the octahedral (O$_h$) point group 
symmetry which permits the following classification for the gap functions:
\begin{eqnarray}
A_{1g}: \psi &\sim& 1, {\rm \ or \ } \left(x^2+y^2+z^2 \right) \nonumber \\
A_{2g}: \psi &\sim& (x^2-y^2)(y^2-z^2)(z^2-x^2) \nonumber \\
E_{g}: \psi &\sim& \left\{ (2z^2-x^2-y^2), \sqrt{3}(x^2-y^2)\right\} \nonumber \\
T_{1g}: \psi & \sim& \left\{xy(x^2-y^2),yz(y^2-z^2),zx(z^2-x^2) \right\} \nonumber \\ 
T_{2g}: \psi & \sim& \left\{xy,yz,zx \right\} \nonumber \\ 
\end{eqnarray}
for the singlet gap functions and 
\begin{eqnarray}
A_{1u}: \psi & \sim& xyz \nonumber \\
T_{1u}: \psi & \sim& \left\{x,y,z \right\} \nonumber \\ 
T_{2u}: \psi & \sim& \left\{x(y^2-z^2),y(z^2-x^2),z(x^2-y^2) \right\} \nonumber \\ 
\end{eqnarray}
for the triplet states.  
\begin{figure}
\includegraphics[width=3.0in]{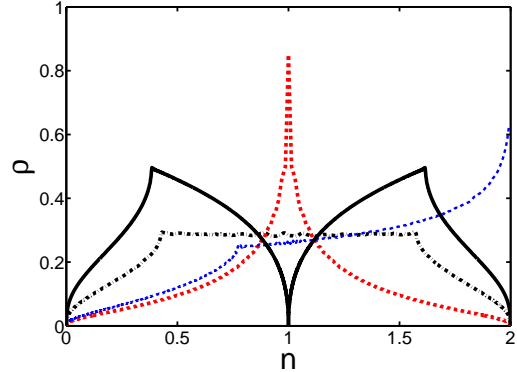}
\caption{Density of states as a function of electron concentration for the diamond  
lattice (solid black curve), SC (dashed-dotted curve), BCC(dashed curve) and FCC (blue dashed curve) lattices.  The FCC is a non-bipartite lattice and therefore, the particle-hole symmetry is absent.      }
\label{dos3d}
\end{figure} 
Figure \ref{dos3d} shows the density of states as a function of electron concentration on each of 
these lattice systems.  With the exception of the BCC and FCC lattices, the density of states remains finite across van Hove singularities in 3 dimensional systems.  

\begin{figure}
\includegraphics[width=3.5in]{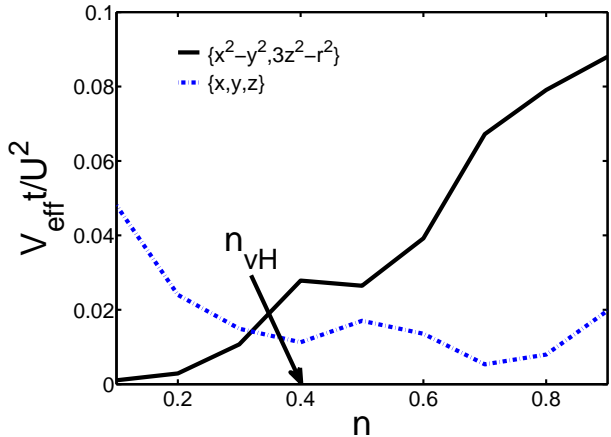}
\caption{Pairing strengths of the dominant pairing configurations on the simple cubic lattice.  
For electron concentrations that are far from half-filling, the 3-fold degenerate p-wave (T$_{1u}$) state occurs 
whereas closer to half-filling the doubly-degenerate  d-wave state (E$_g$) is found.      }
\label{sctc}
\end{figure}

\begin{figure}
\includegraphics[width=3.5in]{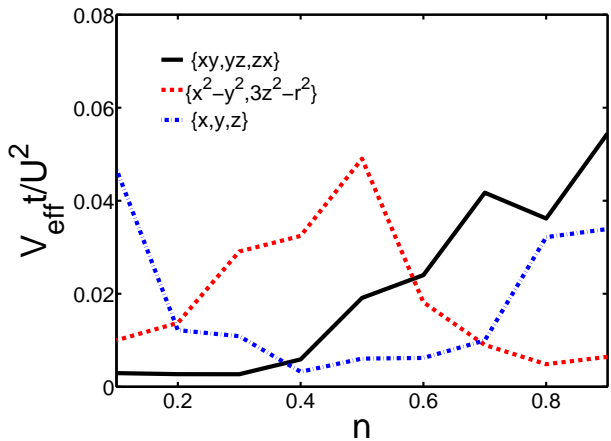}
\caption{Pairing strengths for the dominant pairing configurations on the BCC lattice.  
For electron concentrations that are far from half-filling, the 3-fold degenerate p-wave (T$_{1u}$) state has the greatest strength.  For intermediate concentrations, the E$_g$ and T$_{2g}$ d-wave 
solutions are found to have the strongest strengths.     
      }
\label{bcctc}
\end{figure}

Figures \ref{sctc} and \ref{bcctc} display the phase diagram on the SC and BCC lattice respectively  as a 
function of electron concentration.  At low concentrations, the p-wave solution has the highest 
T$_c$ in both lattices whereas near half-filling, a d-wave solution has the higher transition temperature.  
On the SC lattice the E$_g$ d-wave configuration is found near half filling.  This state is the 3 dimensional analog of the $d_{x^2-y^2}$ gap found on the square lattice near half-filling.  Due 
to the underlying symmetry of the cubic lattice, the $d_{x^2-y^2}$ gap must be degenerate with 
the $d_{z^2-r^2}$ gap function.  This degeneracy in turn implies that below T$_c$, the system 
on the cubic lattice will spontaneously break time-reversal symmetry and form a $d+id$ gap function.  
On the BCC lattice (Fig. \ref{bcctc}), the dominant pairing configuration near half-filling is the triply degenerate d-wave gap function (T$_{2g}$).  At intermediate concentrations, the E$_g$ gap functions 
have the largest T$_c$ and again at low concentrations, the p-wave solution is favored.  
\begin{figure}
\includegraphics[width=3.5in]{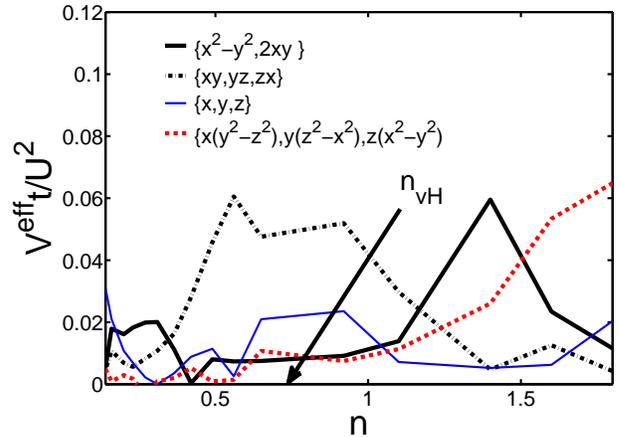}
\caption{ Pairing strengths for electrons on a FCC lattice.  As in the other 3D lattice systems, 
p-wave superconductivity occurs for low electron concentrations.  For intermediate concentrations, the 
 d-wave solutions have the largest pairing strengths.  Near the top of the band, the density 
of states diverges and a 3-component f-wave superconductor has the strongest pairing strength.    }
\label{fcctc}
\end{figure} 
\begin{figure}
\includegraphics[width=3.5in]{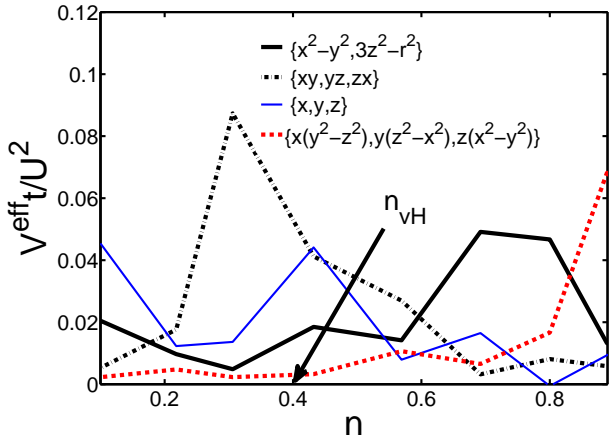}
\caption{ Pairing strengths for electrons on a diamond lattice.  For low electron concentrations, the p-wave solution has the largest pairing strength whereas near half-filling, 
the triply degenerate f-wave gap  has the greatest pairing strength.  For intermediate concentrations, 
both the doubly and triply degenerate d-wave pairing states have the dominant pairing strengths.    }
\label{diamondtc}
\end{figure} 
On the FCC lattice (Fig. \ref{fcctc}) the T$_{2u}$ 
f-wave gap is favored near the top of the band.   This f-wave state gives way to the E$_g$ d-wave state below $n=1.5$, which in turn is replaced by the T$_{2g}$ d-wave configuration below $n=1.3$.  
This triply degenerate d-wave gap function persists for a wide range of concentrations.  Finally, at a 
concentration below $n=0.1$ we find the p-wave state has the highest pairing strength.  
Finally, on the diamond lattice (Fig. \ref{diamondtc}), the T$_{2u}$ gap occurs near half filling when the semi-metal 
is lightly doped.   As the concentration is increased, the ground state consists first of the  E$_g$ and subsequently the T$_{2g}$ superconducting states.  Near the top and bottom of the band, the p-wave solution has the largest pairing strength.

 \section{RG strategy}

 At any given temperature, the thermodynamic properties of the model can be computed perturbatively in powers of $U/t$ so long as $U/t$ is small compared to a characteristic
 $T$ dependent magnitude.  (Indeed, we generally expect that perturbation theory is convergent at finite temperature, with a finite radius of convergence.  Dynamical properties of the system often depend non-analytically on the strength of the interactions, even at elevated temperatures, but as long as we stick to thermodynamic quantities, this issue should not arise.)
Alternatively, even at $T=0$, if we introduce an artificial low energy cutoff, $\Omega_0$, in the spectrum, 
low order perturbation theory 
is reliable so long as $\Omega > \Omega_{PT}$, where $\Omega_{PT}$ can be obtained by looking at the most divergent terms in each order of perturbation theory, the familiar particle-particle ladders:
\ba
&&\rho |U| \log[W/\Omega_{PT}] = 1;\nonumber \\
&&\Omega_{PT} = W \exp[-1/\rho(E_F)|U|].
\ea
$\Omega_{PT}$ is a physical energy scale in the problem - it is the highest energy at which the bare interactions begin to be significantly renormalized by many-body effects.

With this in mind, we formulate the problem in terms of a Grassman path integral, which we express in terms of the normal modes of the quadratic piece of the action defined by $H_0$.  As a first step, we integrate out all the modes with energies greater than a cutoff, $\Omega_0$, chosen so that 
\begin{equation}
U^2/t \gg \Omega_0 \gg \Omega_{PT} .
\end{equation}
Because $\Omega_0 \gg \Omega_{PT}$, the interactions in the resulting effective action can be computed using straightforward perturbation theory.  Moreover, we are guaranteed that in dimensionless units, all the effective interactions are still weak.
Because $W \gg U^2/t$, the 
resulting effective action 
involves 
only modes within a parametrically narrow window, of width $\Omega_0$, about the Fermi surface.  In particular, this effective action is of precisely the form assumed as the starting point for the perturbative RG analysis of the Fermi liquid.\cite{Shankar1994,Polchinski1992}  Specifically,  the dispersion can be linearized about the Fermi surface, the effects of small irrelevant terms can be neglected, and the beta function for the marginally relevant interactions can be computed
 to one loop order.

The second step is to compute the RG flows starting from the initial data obtained in the first perturbative step.  These flows describe how the effective couplings change as we continue the process of integrating out high energy modes by reducing the cutoff below $\Omega_0$.  These equations cease to be accurately governed by the perturbative beta function when one or more dimensionless interaction grows to be of order 1.  However, the value of the cutoff, $\Omega^\star$, at this point defines (up to a multiplicative constant of order 1), a characteristic energy scale in the problem.  Assuming that a one parameter scaling theory describes the low energy physics, then all emergent energies in the problem, including $T_c$, the root-mean-squared gap magnitude, $\Delta_0$, etc., are all simply proportional to $\Omega^\star$.  (Without knowing more explicitly the crossover behavior from the Fermi liquid to the superconducting fixed point, it is not possible to obtain a precise value for the constants of order 1, and hence $\alpha_0$ in Eq. \ref{Tc} cannot be computed by the present methods.)

Note that $\Omega_0$ in this treatment is not a physical energy scale, but rather a calculational convenience.  It is important that the results should be independent of the value of $\Omega_0$.  We will see that by chosing $\Omega_0 \ll U^2/t$, we make simple the manipulations that insure that our results are independent of $\Omega_0$, at least to the desired order in powers of $U/t$.  This highlights an important difference between the present analysis and a conceptually similar treatment of the small $U$ problem considered by Zanchi and Schulz\cite{Zanchi1996}, in which a two step RG analysis of the Hubbard model was undertaken, but in which the intermediate scale (which they call $\ell_x$) is a physical  $U$ independent measure of the proximity to a van-Hove point.  (We return to this comparison in Sec. \ref{relation}, below.)

 \section{First stage renormalization:  perturbative results}
 
 \begin{figure}
\includegraphics[width=3.0in, angle = -90]{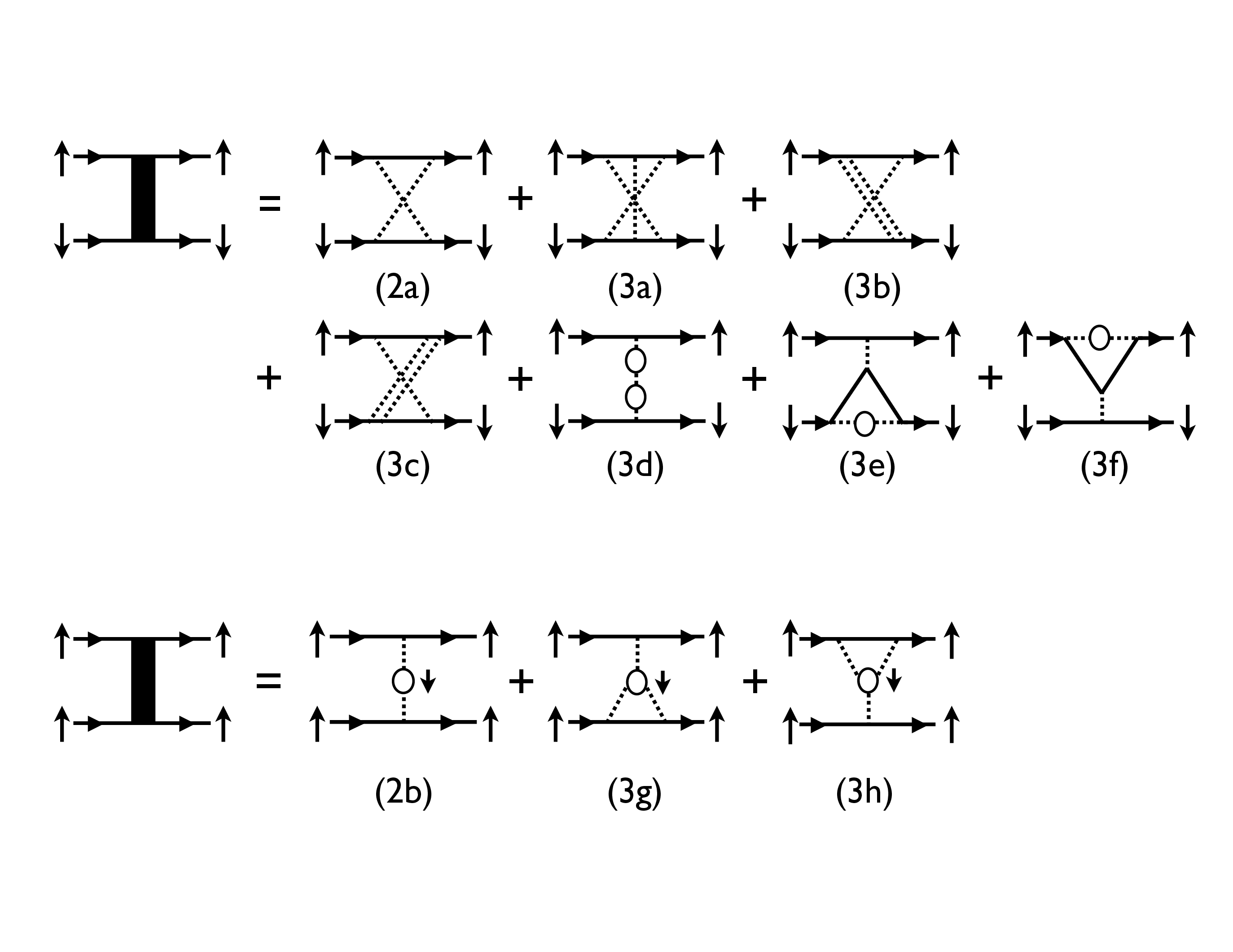}
\caption{ Two particle irreducible diagrams to third order in $U$ which contribute to the effective interaction in the particle-particle 
channel.   The incoming set of electrons at the left of each diagram have  momentum $\vec{k}, -\vec{k}$ and are 
scattered by the interaction to states with momenta $\vec{q}, -\vec{q}$.   
The dashed line denotes the bare Hubbard interaction, and the solid lines correspond to the bare electron 
propagators.  The diagrams are constrained by the requirement that the spin must be flipped across 
the dashed line.} 
\label{irreducible}
\end{figure} 
\begin{figure}
\includegraphics[width=2.5in, angle=-90]{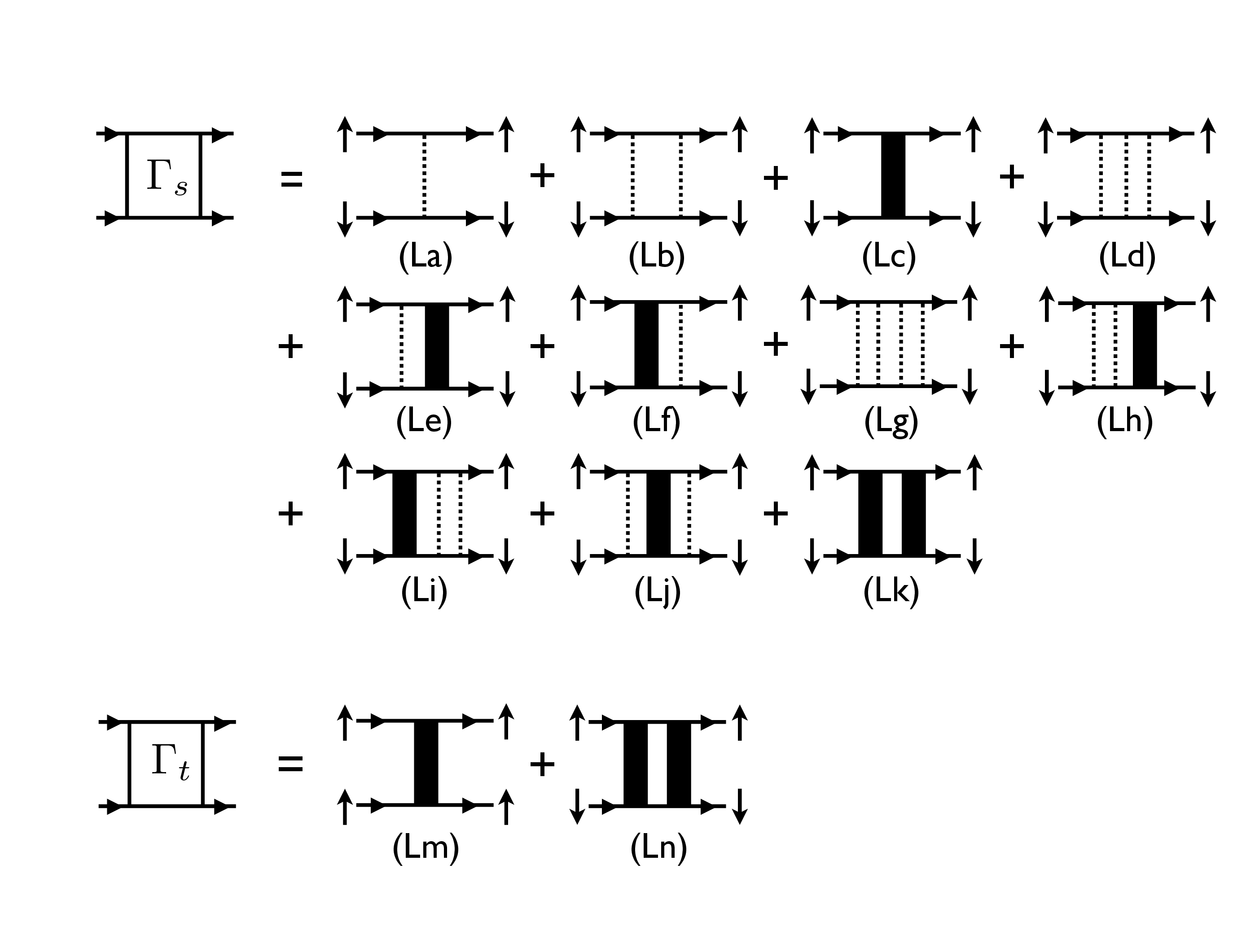}
\caption{Diagrams to $\mathcal{O}(U^4)$ which are used in determining the effective 
interaction in the particle-particle channel.  The upper set of diagrams (L$_a$-L$_k$) show the contributions in the spin singlet channel, whereas the lower set (L$_m$, L$_n$) show the processes that contribute in the triplet channel.     }
\label{ladder}
\end{figure} 
The first step is to integrate out the states with energies down to  $\Omega_0$, and to compute the resulting effective interactions perturbatively in powers of  $U$.   The important terms, which will serve as inputs for the second stage renormalization, are the electron self energy, $\Sigma(\vec k)$, and the two particle vertex, $\Gamma_{\sigma,\sigma^\prime}(\vec k,\vec q)$, in the particle-particle channel i.e. the scattering amplitude of a  pair of particles with spin polarization $\sigma$ and $\sigma^\prime$ and momenta $\vec k$ and $-\vec k$ into a pair of particles with the same spin polarization and momenta $\vec q$ and $-\vec q$.  The two particle vertex can  be decomposed into the singlet and triplet 
channels, $\Gamma_s$ and $\Gamma_t$ respectively, where 
\begin{eqnarray}
\Gamma_{s  }(\vec{k},\vec{q}) &=&  \frac{1}{2} \left[ \Gamma_{\uparrow, \downarrow}(\vec{k}, \vec{q})  
+ \Gamma_{\downarrow, \uparrow}(\vec{k}, \vec{q})  \right] \nonumber \\
\Gamma_t(\vec{k}, \vec{q}) &=& \Gamma_{\uparrow, \uparrow}(\vec{k}, \vec{q})
\end{eqnarray}
The electron self-energy and two-particle vertices are 
\begin{eqnarray}
\Sigma(\vec k;\Omega_0) &=& U^2 \Sigma_2(\vec k;\Omega_0)+ {\cal O} (U^3) \\
\label{Sigma}
\Sigma^{(2)}(\vec k;\Omega_0) &=&\int_{\Omega_0} \frac {d\vec q\  d\omega}{(2\pi)^{d+1}} G(\vec k+\vec q,\omega) G(\vec q,\omega)  \\
&\equiv&   \chi(\vec k;\Omega_0) 
\nonumber 
\end{eqnarray}
\begin{eqnarray}
\Gamma_s(\vec{k}, \vec{q} ; \Omega_0) &=& U + \sum_{n\ge 2}U^n \Gamma_s^{(n)}(\vec{k}, \vec{q}; \Omega_0)  \\
\Gamma_t(\vec{k}, \vec{q}; \Omega_0) &=& \sum_{n \ge 2} U^n \Gamma_t^{(n)}(\vec{k}, \vec{q}; \Omega_0) \\
\label{Gamma}
\Gamma_s^{(2)}(\vec k,\vec q;\Omega_0) &=& \chi(\vec k+ \vec q;\Omega_0) +P(\Omega_0) \label{Gamma2s} \\
\Gamma_t^{(2)}(\vec{k}, \vec{q} ; \Omega_0) &=& - \chi(\vec{k} - \vec{q}; \Omega_0) \label{Gamma2t} \\
P(\Omega_0)&=&\int_{\Omega_0} \frac {d^d p\  d\omega}{(2\pi)^{d+1}} G(\vec p,\omega) G(-\vec p,-\omega)
\end{eqnarray}
Here, $G(\vec k;\omega)$ is the  single-particle Green function, $\int_{\Omega_0}$ signifies the integral over all $\vec q$ subject to the constraints $|\epsilon_{\vec q}|>\Omega_0$, and $|\epsilon_{\vec k+\vec q}| > \Omega_0$, and $d$ is the number of spatial dimensions.  The first term in Eq. \ref{Gamma2s} is obtained from diagram (2a) of Fig. \ref{irreducible} whereas the second term is obtained from diagram (L$_b$) in Fig. \ref{ladder}.  The quantity in Eq. \ref{Gamma2t} is obtained from diagram (2b) in Fig. \ref{irreducible}.   
Below, we shall 
discuss the higher order terms, $\Gamma^{(n)}$ with $n > 2$.
 
 The particle-hole bubble, $\chi$, is regular in the limit $\Omega_0\to 0$, while the particle-particle bubble has the well known logarithmic divergence associated with the Cooper instability, but is otherwise regular:
\begin{eqnarray}
  \chi(\vec k;\Omega_0)=&& \chi(\vec k) + {\cal O} (\Omega_0) \nonumber \\
P(\Omega_0)=&&\rho\log[A/\Omega_0]+ {\cal O} (\Omega_0)
\label{log}
\end{eqnarray}
where 
$A$ is determined by the band-structure over the entire band.  
Since we will never need to keep terms higher order than $U^4$, and since, by assumption, $\Omega_0 \ll U^2$, the higher order terms in powers of $\Omega_0$ can be neglected henceforth.

The higher order vertex functions can likewise be evaluated  by keeping a non-zero value  of $\Omega_0$ where ever there is a logarithmic divergence in the $\Omega_0\to 0$ limit, but setting $\Omega_0=0$ elsewhere.  The expression in the singlet channel is 
\begin{widetext}
\begin{eqnarray}
\Gamma_{s}^{(3)}(\vec k,\vec q;\Omega_0) = && \rho^2\log^2[A/\Omega_0]
+ [\gamma^{(3)}(\vec k) + \gamma^{(3)}(\vec q)] \rho\log[A/\Omega_0] 
 + \tilde\Gamma_{s}^{(3)}(\vec k,\vec q)+ \mathcal{O}(\Omega_0)
\label{Gamma3s}
\end{eqnarray}
\end{widetext}
where if we define $\hat p$ to designate a vector on the (unperturbed) Fermi surface and $S_F\equiv\int d\hat p_F$ equal to the ``area'' of the Fermi surface 
\begin{equation}
\gamma^{(3)}(\vec k) \equiv \int \frac {d\hat p}{S_F}\left(\frac {\bar v_F}{v_F(\hat p)}\right) \chi(\vec k + \hat p)
\end{equation}
with $v_F(\hat p)$ the magnitude of the Fermi velocity at position $\hat p$ on the Fermi surface and the norm of the Fermi velocity defined according to
\begin{equation}
\frac 1 {\bar v_F} \equiv \int \frac {d\hat p}{S_F}\left(\frac {1}{v_F(\hat p)}\right).
\end{equation}
The first term of Eq. \ref{Gamma3s} is obtained from  diagram (L$_d$) in Fig. \ref{ladder} whereas the terms proportional to $\log \left[ A/\Omega_0 \right]$ are derived from diagrams (L$_e$,L$_f$) in Fig. \ref{ladder} with the thick solid line treated to second order in U.    Finally, $\tilde\Gamma^{(3)}$ contains all the non-singular contributions in the limit $\Omega_0\to 0$, which are 
derived from the third-order two-particle-irreducible (2pI) diagrams (3a-3f in Fig. \ref{irreducible}).  
They can be expressed as double momentum integrals over suitable products of quartets of $G$'s, 
as shown in the appendix, but in
the interest of 
clarity, we do not display them here.
In the 
triplet channel, the third-order correction to the vertex is non-singular and is obtained from diagram (3g) of 
Fig.
\ref{irreducible}: 
\begin{equation}
\Gamma_{t}^{(3)}(\vec k,\vec q;\Omega_0) = \tilde{\Gamma}_t^{(3)}(\vec{k}, \vec{q}) + \mathcal{O}(\Omega_0)
\end{equation}
An explicit expression for this quantity is also derived in the appendix.  

Similarly, we can obtain an expression for $\Gamma^{(4)}$:
\begin{widetext}
\begin{eqnarray}
\Gamma_{s}^{(4)}(\vec k,\vec q;\Omega_0) = &&\Big\{\rho^3\log^3[A/\Omega_0]
+ [\gamma_1^{(4)} + \gamma_1^{(3)}(\vec k) + \gamma_1^{(3)}(\vec q)] \rho^2\log^2[A/\Omega_0]\nonumber \\
&& + [\gamma_2^{(4)}(\vec k) + \gamma_2^{(4)}(\vec q)] \rho\log[A/\Omega_0]
+\gamma_s^{(4)}(\vec k,\vec q)\rho\log[A/\Omega_0]\Big\}  +  \tilde\Gamma_{s}^{(4)}(\vec k,\vec q)+\mathcal{O}(\Omega_0)
\label{Gamma4}
\end{eqnarray}
\end{widetext}
where
\begin{eqnarray}
\gamma_1^{(4)} &=& \int \frac{d \hat{p} d \hat{p}'}{S_F^2} \frac{\bar{v}^2_F}{v_F(\hat{p}) v_F(\hat{p}')}
\chi(\hat{p} + \hat{p}') \\
\gamma_2^{(4)}(\vec{k}) &=&  \int \frac {d\hat p}{S_F}\left(\frac {\bar v_F}{v_F(\hat p)}\right) \tilde{ \Gamma}^{(3)}_s(\vec{k}, \vec{p})
 \\
\gamma_s^{(4)}(\vec k,\vec q) &=& \int \frac {d\hat p} {S_F}
\chi(\vec k +\hat p)  \left(\frac{\bar v_F}{v_F(\hat p)}\right) \chi(\vec q + \hat p)
\end{eqnarray}
The first term in Eq. \ref{Gamma4} is represented by the fourth order ladder diagram in (L$_g$) of Fig. \ref{ladder}.  The  terms involving $\gamma_1^{(n)}$ are obtained from diagrams (L$_h$-L$_j$), and those involving $\gamma_2^{(4)}$ are derived from diagrams (L$_e$,L$_f$), treating the thick solid line to 3rd order in U.  
Finally, $\tilde{\Gamma}_s^{(4)}$ is a non-singular contribution from the fourth-order 2PI diagrams, which are not shown here 
Since we shall not make use of these terms, we will not 
provide explicit expressions for them.  In the triplet channel, the fourth-order vertex consists of a single term
\begin{equation}
\Gamma_t^{(4)}(\vec{k}, \vec{q}; \Omega_0) = \rho \log[A/\Omega_0] \gamma_t^{(4)}(\vec{k}, \vec{q})+
\mathcal{O}(\Omega_0)
\end{equation}
where
\begin{equation}
\gamma_t^{(4)}(\vec{k}, \vec{q}) = \int \frac {d\hat p} {S_F}
\chi(\vec k - \hat p)  \left(\frac{\bar v_F}{v_F(\hat p)}\right) \chi(\vec p - \hat q)
\end{equation}
which is obtained from diagram (L$_n$) of Fig. \ref{ladder}, treating the thick solid line to second order in U.  
 \section{Second stage results:  RG analysis}
 
 The second stage of renormalization is carried out following the standard Fermi liquid renormalization group (RG) procedure of Shankar\cite{Shankar1994} and Polchinski\cite{Polchinski1992}.  Notice that the effective action generated after the first stage of renormalization is precisely of the form assumed as the starting point of this renormalization procedure:  the effective interactions are all small (so a perturbative RG approach is justified) and the remaining states lie in a narrow strip of width $\Omega_0$ about the Fermi surface so that the spectrum can be linearized without loss of accuracy.  Various interactions (Fermi liquid parameters) are marginal at the non-interacting fixed point.  These do not significantly affect our principle results.   The only couplings that renormalize are those in the Cooper channel.  These are governed by the one-loop RG equations which we write in matrix form as
 \begin{equation}
\frac {dg}{d\ell} = -g \star g 
\label{RG}
\end{equation}
where $\ell\equiv \log[\Omega_0/\Omega]$,
\begin{equation}
(g \star h)_{\vec k,\vec q}
\equiv \int \frac {d\hat p}{S_F} g_{\hat k,\hat p} h_{\hat p,\hat q},
\end{equation}
where $g$ designates a dimensionless matrix 
\begin{equation}
g_{\hat k,\hat q} \equiv \rho \sqrt{ \frac {\bar v_F} {v_F(\hat k)} } \Gamma
(\hat k,\vec q) \sqrt{\frac{\bar v_F} {v_F(\hat q)} }.
\label{g}
\end{equation}
Here, we have left implicit the dependence of both $g$ and $\Gamma$ on the spin indices.       

In integrating these equations, we start with an initial value of the interaction matrix, $g^0$, which is the output of the first stage of renormalization.  Because $g$ is a real symmetric matrix, it is also Hermetian, and so can be diagonalized:
\begin{equation}
g^0_{\vec k,\vec q} = \sum_n \lambda_n^0 \psi_n^\star(\vec k)\psi_n(\vec q)
\end{equation}
where the eigenvalues, $\lambda_n$, are real and the eigenfunctions form an orthonomral basis,
\begin{equation}
\int \frac {d\hat k}{S_F} \psi_n^\star(\vec k)\psi_m(\vec k) = \delta_{n,m}.
\end{equation}
As a result, each eigenvalue renormalizes independently:
\begin{equation}
\frac {d\lambda_n}{d\ell} = -\lambda_n^2; \ \ \ \lambda_n(\Omega) =\frac {\lambda_n^0} 
{1+\lambda_n^0\log[\Omega_0/\Omega]}.
\end{equation}

To determine the physically important scale, $\Omega^\star$, at which the RG treatment breaks down, we must first identify the smallest eigenvalue of $g^0$, $\lambda_0^0$, for which  $\lambda_0^0\leq \lambda_n^0$ for all $n > 0$.  Assuming that $\lambda_0^0$ is negative, then 
\begin{equation}
\Omega^\star = \Omega_0\exp[-1/|\lambda_0^0|]
\label{Omegastar}
\end{equation}
Asymptotically, both $T_c$ and the zero temperature gap scale, $\Delta_0$, are then equal to a (unknown) number of order 1 times $\Omega^\star$.
Under generic circumstances, we expect $\lambda_0$ to be non-degnerate, {\it i.e.} $\lambda_0 < \lambda_1$.  The exception to this is the vicinity of a zero temperature phase transition between two superconducting states with different symmetries, where two eigenvalues cross.  The properties of the infinitesimal neighborhood of such critical points will not be investigated further in this paper.

\begin{figure}
\includegraphics[width=2.in]{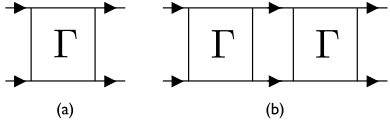}
\caption{ An arbitrary irreducible interaction vertex $\Gamma(\vec{k}, \vec{q})$  in the particle-particle channel is 
shown in (a).   The spin indices are suppressed for clarity.  
The flow of the most negative eigenvalue of $\Gamma$ breaks down at a scale which is identified as the superconducting transition temperature.  For each such vertex, there belongs a corresponding  diagram of the form shown in (b), which produces the logarithmic divergence $ \sim \log (A/ \Omega_0)$ needed 
to remove the dependence on the arbitrarily chosen initial cutoff $\Omega_0$. }
\label{cutoff}
\end{figure} 

There is an apparent problem with Eq. \ref{Omegastar}, which is that it has an explicit dependence on $\Omega_0$.  Since $\Omega_0$ was introduced as an unphysical calculational device, this dependence must be spurious.  Fortunately, there is also an implicit dependence of $\lambda_0^0$ on $\Omega_0$, which just cancels this explicit dependence.  

To see this most simply, first consider the case of the negative $U$ Hubbard model.  In this case, the lowest eignevalue is clearly in the spin singlet channel, and it can be computed perturbatively in powers of $U$ as
\begin{eqnarray}
\psi_0(\vec k) =&& \sqrt{\frac{v_F(\hat k)}{\bar v_F} }[ 1 + {\cal O}(U)] \nonumber \\
\lambda_0^0 = &&\rho U + \rho U^2 [ P(\Omega_0) +  \rho' ]+{\cal O}(U^3)
\label{negU}
\end{eqnarray}
where
\begin{equation}
\rho'=\int \frac{d\hat k d\hat q}{S_F^2} \chi(\hat k+\hat q).
\end{equation}
The logarithmic dependence of $P$ on $\Omega_0$ insures that  when the expression in Eq. \ref{negU} is inserted into Eq. \ref{Omegastar}, the result is independent of $\Omega_0$, at least to the stated order in powers of $U$:
\begin{equation}
T_c \sim \Omega^\star=A\exp[-(1/\rho|U|) e^{-\rho'/\rho}\Big [ 1 + {\cal O}(U)\Big ].
\end{equation}


The same analysis can be carried through for the more complicated case of the repuslive $U$ Hubbard model.  In this case, however, since the leading contribution to $\lambda_0^0$ is order $U^2$, 
the term responsible for canceling the $\Omega_0$ dependence of the prefactor is a logarithmically divergent fourth-order order contribution to  
the `dressed' vertex.  The dressed vertex is represented by the thick solid line in diagram (L$_c$) or (L$_m$) of Fig. \ref{ladder}.  We shall consider the singlet and triplet channels separately in what follows.  


In order to see that 
$\Omega^*$  in the singlet channel does 
 not depend on $\Omega_0$, it is helpful to discretize the points on the Fermi surface, so that the 
matrix, $\Gamma_s$ in Eq. \ref{Gamma}, and hence the matrix, $g_s$ in Eq. \ref{g}, as well, 
is a $N \times N$ matrix, where $N$ is the number 
of 
k-points.
(The continuum limit can easily be taken at the end of the calculation.) 
  This $N\times N$ matrix can then be partitioned into a $1 \times 1$ block $g_{s,0}$ which affects only the 
trivial s-wave states, and an $\left(N-1 \right) \times \left( N-1 \right)$ block $g_{s,1}$ containing the non-trivial superconducting states: 
\begin{equation}
g_s(\hat{k}_i, \hat{k}_j) = 
\left(\begin{array}{cc}
g_{s,0} &  \mathcal{T}_s  \\
\mathcal{T}^{\dagger}_s & 
g_{s,1} \\
\end{array}\right) 
\end{equation}
Here, $\mathcal{T}_s$ is a  $1 \times (N-1)$ matrix which connects the trivial s-wave subspace with 
the orthogonal space of sign-changing pair-fields.  

The lowest order contributions to $g_{s,0}\propto\Gamma_{s,0}$ comes from diagrams (L$_a$,L$_b$) in Fig. \ref{ladder} 
so $\Gamma_{s,0} \sim \mathcal{O}(U)$. 
On the otherhand, $g_{s,1}\propto\Gamma_{s,1}\sim \mathcal{O}(U^2)$.  
The diagrams which contribute to $\mathcal{T}_s(\hat{k}_i, \hat{k}_j)$ are shown in (L$_e$) and (L$_h$) in Fig. \ref{ladder}.  The interaction vertex in these diagrams has the form of a product of two terms, one of which consists of a bare (and therefore momentum-independent) vertex operating on the incoming momenta 
$(\hat{k}_i)$, and the other of which is a dressed vertex connected to the outgoing momenta $(\hat{k}_j)$.  Therefore, these 
terms, when viewed as a matrix, operate on a non-trivial singlet pairing configuration and produce the 
trivial s-wave solution.  From Figs. \ref{irreducible} and \ref{ladder}, it is easy to see that $\mathcal{T}_s \sim \mathcal{O}(U^3)$.  In a similar fashion, diagrams (L$_f$) and (L$_j$) contribute to $\mathcal{T}^{\dagger}_s$ which is just 
the hermitian conjugate of $\mathcal{T}_s$.  
In diagonalizing 
$g_s$, the off-diagonal matrix elements, $\mathcal{T}_s$, can be treated perturbatively.  Indeed, it is clear that the leading effect of these terms on the eigenvalue problem is ${\cal O}(U^5)$, and so to the order we are working, they can be set to zero.

Within the non-trivial $(N-1)$-dimensional subspace, the lowest-order term in perturbation theory which contribute to 
$\Gamma_{s,1}$ is represented by the dressed vertex in diagram (L$_c$) of Fig. \ref{ladder}.  To lowest 
order in $U/t$, this is the diagram in (2a), which is $U^2 \chi(\hat{k}_i + \hat{k}_j ; \Omega_0)$, and 
when all of diagrams (2a)-(3g) are taken into account, the non-singular terms of $\mathcal{O}(U^3)$ 
which produce $\tilde{\Gamma}^{(3)}_s(\hat{k}_i, \hat{k}_j)$ in Eq. \ref{Gamma3s} will also contribute to the vertex.  
We define the quantity $\tilde \Gamma_{s,1}$ to be the piece of $\Gamma_{s,1}$ which is non-singular in the limit $\Omega_0\to 0$ 
\begin{equation}
\label{Gammaprime}
\tilde\Gamma_{s, 1}(\hat{k}_i, \hat{k}_j) = U^2 \chi(\hat{k}_i + \hat{k}_j) + U^3 \tilde{\Gamma}^{(3)}_s(\hat{k}_i, \hat{k}_j)+{\cal O}(U^4)
\end{equation}
and  correspondingly
\be 
\tilde g_{s,1}(\hat k,\hat q)=\rho\sqrt{\bar v_F/v(\hat k)}\ \tilde \Gamma_{s,1}(\hat k,\hat q)\ \sqrt{\bar v_F/v(\hat q)}.
\ee

The lowest order term in 
perturbation theory for $\Gamma$ which has singular (logarithmic) $\Omega_0$ dependence is the same diagram (L$_b$) in Fig. \ref{ladder} that gives the logarithm in Eq. \ref{negU}; however, this term is purely a contribution to $\Gamma_{s,0}$.  The lowest order singular term in $\Gamma_{s,1}$ which has the singular $\Omega_0$ dependence 
derives from fourth-order diagram (L$_k$) in Fig. \ref{ladder}.
Thus, when all the diagrams are properly taken into account, 
we see that
\begin{equation}
g_{s,1} = \tilde g_{s,1}
+ \tilde g_{s,1} 
\star\tilde g_{s,1} 
\ \log\left[A/\Omega_0 \right]+ {\cal O}(U^5) 
\end{equation}
Thus, if we express the results in terms of $\tilde\lambda$, the eigenvalues of the non-singular part of the interaction, $\tilde g_{s,1}$, we find that
\begin{equation}
T_c = A\exp{ \left\{- 
1/\vert \tilde \lambda_{min}\vert 
\right\} }  \left[ 1 + {\cal O}(U^2)\right]
\end{equation} 
The logarithmic dependence on $\Omega_0$ of the 
fourth order contribution to $\lambda_0^0$ has just the requisite form to cancel the explicit dependence on $\Omega_0$.  

The eigenvalue obtained this way is valid to $\mathcal{O}(U^3)$.  However, to obtain the quantity $\alpha_2$, one only needs to consider $g_{s,1}$ to $\mathcal{O}(U^2)$, which is obtained from the first term in Eq. \ref{Gammaprime}.  Once the eigenvalue problem to this order has been solved, the $\mathcal{O}(U^3)$ correction, which we refer to as $\alpha_1$, is obtained by treating the contribution from the second 
term in Eq. \ref{Gammaprime} to first order in perturbation theory.  

It is similarly straightforward to show that $T_c$ for a spin-triplet ground state is independent 
of the initial cutoff $\Omega_0$.  Again, the effective interaction in the triplet channel must have 
non-trivial momentum dependence, and only the dressed vertex shown in Fig. \ref{irreducible} can 
contribute.  The resulting expression in the triplet channel is directly analogous to Eq. \ref{Gammaprime}:
\begin{equation}
\tilde\Gamma_{t, 1}(\hat{k}_i, \hat{k}_j) = -U^2 \chi(\hat{k}_i - \hat{k}_j) + U^3 \tilde{\Gamma}^{(3)}_t(\hat{k}_i, \hat{k}_j)
+{\cal O}(U^4)
\end{equation}
The dependence of $T_c$ on the initial scale $\Omega_0$ is eliminated by diagram (L$_n$) in 
Fig. \ref{ladder} which possesses the required logarithmic divergence $\sim \log\left[A/\Omega_0\right]$.  After including both diagrams (L$_m$) and (L$_n$) in Fig. \ref{ladder}, we find that the interaction 
vertex in the triplet channel is 
\begin{equation}
g_{t,1} = 
\tilde g_{t,1} 
-\tilde g_{t,1}\star \tilde g_{1,t} \ \log\left[A/\Omega_0 \right] +{\cal O}(U^5)
\end{equation}
Thus, as was found in the singlet channel, the final expression for $T_c$ is independent of 
$\Omega_0$.

The general feature in perturbation theory which acts to eliminate the initial cutoff dependence 
of $T_c$ is shown in Fig. \ref{cutoff}.  In Fig. \ref{cutoff}(a), an arbitrary irreducible interaction vertex $\Gamma$ in the 
particle-particle channel is shown.  For the negative U Hubbard model, $\Gamma$ would correspond 
simply to the bare vertex, and for the repulsive cases, $\Gamma$ is the appropriate irreducible vertex 
in either the singlet or triplet channel as discussed above, and could be computed to an arbitrary 
order in perturbation theory.  For each such $\Gamma$, there corresponds a diagram of the form 
shown in \ref{cutoff}(b) which acts to remove the dependence on $\Omega_0$.  In this diagram, 
the internal legs which separate each vertex $\Gamma$ produces the required logarithmic divergence factor 
$P(\Omega_0)$.  Since this is true for an arbitrary interaction vertex, the fact that the final 
expression for $T_c$ is independent of the arbitrarily chosen initial scale $\Omega_0$ is 
true to all orders in perturbation theory for the interaction vertex (we note in passing that self 
energy corrections to the internal propagator lines produces $\mathcal{O}(U^2)$ corrections   .  
in the Hubbard model and do not affect the general structure discussed here).




\section{Relation to previous work}
\label{relation}

The present work is, in part, 
a recasting of old work in a new framework, in terms of a more well controlled asymptotic analsyis. Specifically, Kohn and Luttinger\cite{Kohn1965} observed that even for a repulsive bare interaction,  the momentum-dependent structure in the irreducible particle-particle vertex can give rise to a Cooper 
instability in a suitable channel.  For a short range repulsive bare interaction, the resulting pairing 
interaction is mediated by an $S=1$ particle-hole channel which is the leading term of the Berk-Schrieffer spin fluctuation exchange interaction\cite{Berk1966}.  


An important piece of the physics of correlated materials that is missing in the weak coupling limit is associated with ``competing  orders,'' and the accompanying interplay between interactions in different ``channels."  We stress that this is not a failure of the method of solution, but is something that is an intrinsic feature of a weakly perturbed Fermi liquid.   In order to address this physics, various calculations have been undertaken using the Functional Renormalization Group (FRG) method\cite{Salmhofer1998,Binz2003,Halboth2000,Honerkamp2001} which is in many ways similar in structure to the weak-coupling analysis undertaken here.  In this approach, a single set of perturbative non-linear flow equations are derived for the coupling constants defined both near to and far from the Fermi surface, and then the resulting flows are analyzed numerically starting from initial conditions corresponding, for instance, to the Hubbard model with intermediate couplings, $U \sim 3 - 4t$.  As stressed in Ref.\cite{Zhai2009}, for these values, the effective interactions near the Fermi surface are 
typically large.  On an intuitive level, the advantage of the FRG approach is that it does capture some physics of multiple intertwined scattering processes in a physically compelling manner.  On the other hand, the perturbative methods used are formally valid only in the asymptotic limit $U/t \to 0$.  While, as far as we know, there has been no published work analyzing the FRG flows for the Hubbard model in this limit, were the FRG calculations carried out in this limit they would approximate the leading order behavior derived in the present paper using somewhat different methods.

Another feature of the weak coupling limit is that the dynamics is set by the bare bandwidth. 
Thus, the bare susceptibilities that enter the expression for the pairing interaction are those of the unperturbed Fermi gas, in which the only energy scale is the bandwidth.  Again, a plausible extension of the present results involves replacing one or more of these susceptibilities with a dressed susceptibility, or even with an experimentally measured susceptibility.  For example, near a magnetically ordered 
phase, such a susceptibility would reflect  the enhancement of the magnetic fluctuations for $\vec k \sim \vec Q$, the antiferromagnetic ordering vector, and still more importantly, the retardation effects implicit in the emergence of a new energy scale, ``$J$'', associated with magnetic fluctuations.  From this perspective, the various approaches to a ``spin-fluctuation exchange'' mechanism of superconductivity appear as natural extrapolations of the present results to a more strongly coupled regime.
However, it is important to stress that such procedures represent uncontrolled approximations.  There 
may well be competing channels or possibly a failure of the basic framework.  One would like to have 
numerical calculations to test the utility of such phenomenological approximations.  
There are several specific studies we wish to discuss explicitly, as they have produced results which are particularly close to those obtained here:

The two step RG approach of Schulz\cite{Zanchi1996} has already been mentioned.  In the case considered there, the role of $\Omega_0$ was played by a physical energy scale, defined as the energy scale below which the singular structures due to the proximate van Hove singularity could be ignored, and above which the system might as well be precisely tuned to the van Hove point.  In particular, in that case, $\Omega_0$ (which they call $\ell_x$)\cite{Zanchi1996} takes on a $U$ independent value which depends, instead, on $t$, $t'$, and the chemical potential, $\mu$.  Strictly speaking, unless the parameters are fine-tuned to be parametrically close to the van-Hove point, this physics does not survive the small $U$ limit, as we have shown;  our method includes, already, all effects of arbitrary band-structure both at and away from the Fermi surface.  

\section{Discussion and further directions}
\label{discussion}

The results we have obtained are asymptotically exact in the limit $U/t \to 0$, so long as the conventional RG treatment of the Fermi liquid is valid.  In this limit, $T_c$ tends rapidly to zero, so  the present results cannot be directly  associated with a mechanism of ``high temperature superconductivity.''  Moreover, it is clear that in most materials of interest, the interactions are moderate to strong, so the results cannot be said to have any direct relevance to these materials.  However, some aspects of the results seem worth emphasizing which, when extrapolated to stronger coupling (where controlled calculations are not possible), may give insight into mechanisms of high temperature superconductivity in real materials.

Most importantly, we have shown that repulsive interactions combined with lattice induced band-structure effects generically {\em do} result in a superconducting ground-state with non-trivial transformation properties with respect to the point group symmetries of the crystal.  

1)  Where band-structure effects are weak, {\it i.e.} where the Fermi surface is nearly circular or spherical, the dominant superconducting instability is, generically, a two or three-fold degenerate spin-triplet p-wave.  The driving force for this is more or less the same as originally envisaged by Kohn and Luttinger.  However, here if we compare $T_c$ with fixed small $U/t$ at different band-fillings, the values of $T_c$ found in this regime are small compared to the (still small) values obtained where the Fermi surface is more structured.  Since in all the cases we studied, the p-wave state is a 2 or 3 fold degenerate representation of the point group, and the $d$ vector is also arbitrary, there are many different possible forms of ordered state possible in principle.  However, general considerations  suggest that the generic ground-state will either break time-reversal symmetry, forming a  ``topological'' p+ip superconductor (which corresponds to the A-phase of Helium-3 and presumably is the state that is observed in Sr$_2$RuO$_4$ \cite{Mackenzie2003}), or the system will spontaneously break the 
relative ``spin-orbit" symmetry as is the case in the B-phase of Helium-3.  The perturbation which 
lifts the degeneracy between these two possibilities is the spin-orbit coupling of the normal state.  
Thus, an interesting extension of our work would be to incorporate the effects of spin-orbit coupling into the present analysis. 

2)  On the square or tetragonal lattice with $n$ near 1, higher values of $T_c$ are obtained in a spin singlet d-wave channel.  This is encouragingly similar to what is found in the cuprate high temperature superconductors. 

3)  On the triangular, hexagonal, cubic, FCC, and BCC lattices, there is a range of electron concentrations for which the dominant superconducting state is a two or three-fold degenerate singlet d-wave state.  General arguments suggest that, in this case, the superconducting state will be a time-reversal symmetry breaking d+id superconducting state.  While we do not know of a currently well characterized material in which such a superconducting state has been observed, the state is  analogous to the ``anyon  superconducting state" originally proposed by Laughlin\cite{Rokshar1993,Laughlin1994}.  

4)  On both the triangular and hexagonal lattices, when there is more than one electron or hole pocket, the superconducting ground-state is a non-degenerate spin triplet state, in which there is a full gap everywhere on the Fermi surface, but the gap changes sign in going from one pocket to the other.  This situation  gives the highest transition temperatures (for fixed small $U/t$) that we have found.  Moreover, the sign alternation on different pieces of the Fermi surface (although not the triplet character) is reminiscent of the proposed gap function symmetry in the Fe-pnictides. 

Conversely, it is interesting to study what aspects of the physics of real materials are {\em inconsistent} with the behavior of the Hubbard model at small $U/t$.  Taking the cuprate high temperature superconductors as an example, there are several features worth mentioning:  
5)  In the cuprates, $T_c$ is not very small in comparison to  the microscopic energy scales.  Indeed, it reaches values that are, within a factor of two or three, equal to the strong-coupling dimensional analysis estimate $T_c \sim |1-n| J$.  In weak coupling, by contrast, $T_c$ is an emergent energy which is exponentially smaller than any of the microscopic energies in the problem.  (However, $T_c$ does extrapolate to a value of order $t$ in the limit $\rho U \sim 1$, assuming that the extrapolation remains valid even where the justification for the result fails.)
 6)  In the cuprates, there is a clear breakdown of Fermi liquid theory at temperatures above $T_c$, except possibly in the case of the most overdoped materials.  Moreover, there is good evidence\cite{Wang2006,Sondhi2002, Mukerjee2004,Podolsky2007,Raghu2008} of substantial superconducting phase fluctuations persisting well above $T_c$.  In weak coupling, there is an emergent exponentially small crossover scale, $\Omega_{PT}$, at which perturbation theory breaks down, and where a corresponding breakdown of Fermi liquid theory is possible.  However, because $T_c$ is exponentially smaller than $\Omega_{PT}$, Fermi liquid behavior  applies in a wide range of temperatures above $T_c$ --  $\Omega_{PT} \gg T > T_c$.  BCS mean-field theory should, likewise, provide an extremely accurate description of the superconducting transition.  In particular, the characteristic energy scale for phase fluctuations,\cite{Emery1995} $T_{\theta} = (n_s /2m^*) \xi_0^{D-2} \sim E_F (\xi_0/a)^{D-2}$, is exponentially larger than $T_c$, precluding any substantial role for phase fluctuations.  (Here $n_s/2m^*$ and $\xi_0$ are, respectively, the zero temperature superfluid stiffness and coherence length, and $a$ is a lattice constant.)

3)  Direct measurements of the gap function in the cuprates\cite{Shen1993,Ding1995} suggest a gap function with the simple form, $\Delta_{\vec k}= [\cos(k_x) - \cos(k_y)]$, at least in optimally doped materials. In real-space, this form implies the pair-field extends only to nearest-neighbor sites.  In contrast, the weak-coupling gap function is more structured in k-space, as shown in Fig. \ref{wf}.  This reflects the non-local character of the induced pairing interactions and is, we believe, a generic feature of weak coupling.

4)  In the cuprates, superconductivity emerges from a doped antiferromagnetic insulator with well defined spin-wave modes similar to those expected for a spin-1/2 Heisenberg antiferromagnet with exchange coupling $J < t$ (and possibly some higher order exchange coupling representing\cite{Coldea2001} the fact that $U \sim 8t$ is not all that large).  Moreover, there is direct evidence from neutron scattering that spin-wave-like collective excitations of the system with the same characteristic energy scale\cite{Xu2009,Anderson1997} $J$ persist into the superconducting phase at energy scales larger than the gap, even when there is no corresponding broken symmetry.  In weak coupling, an antiferromagnetic insulating state occurs only when two parameters are exponentially fine tuned, $|t^\prime/t| < \delta$ and $|(n-1)| < \delta$, where $\delta \sim \exp[ \ - \alpha\sqrt{t/ U}\ ]$.  A metallic antiferromagnetic state (as well as various other more exotic ordered phases) could occur with the fine tuning of only a single parameter, $|(n-n_{vh})| < \delta$, in the vicinity of a van-Hove singularity.  However, outside of this exponentially narrow range of concentrations, there is no identifiable local antiferromagnetic order present at weak coupling.  Indeed,  without some form of exponential fine-tuning, ``competing orders'' is not a concept that occurs in the weak coupling limit. 

There are many interesting directions in which the present work can be generalized.  

1)  Most straightforwardly, it can apply to the Hubbard model on lattice systems with other band-structures than those considered here.  

2)  It is clearly also possible to extend the same sort of analysis to situations in which there are more complicated interactions than the Hubbard $U$, such as multiband models, where there are both intra-band and interband interactions, and even single band models with longer range interactions, such as a nearest-neighbor repulsion, $V$.  Technically, these interactions complicate the analysis in the sense that a large number of other diagrams, involving the interaction between electrons of like spin, enter the pertubative analysis.  Moreover, there are a variety of ways that the asymptotic analysis can be carried out, which will clearly lead to different physics.  For instance, with interactions $U$ and $V$, the analysis can be carried out as $U/t \to 0$ with $U/V$ held fixed, or with $U/V \sim (U/t)^y$ where $y$ is an appropriate positive exponent.  

3)  Perhaps the most interesting extension involves the fine-tuning discussed above, which may permit the study of various strong-coupling phenomena in the weak coupling limit.  In particular, the physics of competing orders can be explored by performing an asymptotic analysis in the limit $U/t\to 0$ and $n\to n_{vh}$ in such a way that $|n-n_{vh}| \ \exp[\ \alpha \sqrt{1/U}\ ] \to $ constant.  We expect that this will permit us to reproduce much of the interesting physics obtained in various FRG calculations, but in a more controlled fashion.  In order to study the interplay between superconductivity and a Mott insulating (antiferromagnetic) phase, an even more complex analysis must be performed, taking the limit $U/t\to 0$, $n\to 1$, and $t^\prime/t \to 0$ in such a way that $|n-1| \ \exp[\  \alpha \sqrt{1/U}\ ] \to $ constant and $|t^\prime/t| \ \exp[\  \alpha \sqrt{1/U}\ ] \to $ another constant.

\begin{acknowledgments}
We acknowledge helpful discussions with D. Agterberg, E. Berg, A. Chubukov, D. Fisher, E. Fradkin,  P. Hirschfeld, and S. Shenker.  This work was initiated during the KITP workshop on Higher Temperature Superconductivity.  It was supported in part by NSF Grant No. NSF DMR0758356 at Stanford (SAK and SR), and Grant No.  PHY05-51164 at KITP.  DJS acknowledges the Center for Nanophase Materials Science, which is sponsored at Oak Ridge National
Laboratory by the Division of ScientiÞc User Facilities,
U.S. Department of Energy and thanks the Stanford Institute of Theoretical Physics for their hospitality.
\end{acknowledgments}

\appendix*
\section{Perturbation theory in the Cooper channel}
The perturbative corrections to the 2-particle interaction vertices
in the particle-particle channel are shown in Figs. \ref{irreducible} and \ref{ladder}.  
In these diagrams two particles with momentum $\vec{k}$ and $-\vec{k}$ are scattered 
to states with momenta $\vec{q}$ and $- \vec{q}$.  In the singlet channel, the incoming and 
outgoing pair of particles have the opposite spin whereas in the triplet channel, we require them 
to have the same spin.  We first consider the diagrams in the singlet channel.  

The first order contribution to the interaction give in diagram (L$_a$) of Fig. \ref{ladder} is simply 
\begin{equation}
\Gamma_s(L_a) = 1 
\end{equation}
The first order term is independent of momentum and affects only the trivial s-wave pairing states.  Next consider the higher-order terms.  Before proceeding, we introduce as a notational convenience the following definition 
\begin{equation}
\sum_{p} \equiv \int_{\Omega_0} \frac{d  p_0 d^d p }{\left(2 \pi \right)^{d+1}}
\end{equation}
where $p_0$ is a Matsubara frequency.  
  The second-order ladder diagram shown in (L$_b$) of Fig. \ref{ladder} is 
\begin{eqnarray}
\Gamma_s(L_b) &=&  \sum_{p}  G_0(p) G_0(-p) \nonumber \\
&=& \int_{\Omega_0} \frac{d^d p}{\left(2 \pi \right)^{d} } \left[ \frac{1 - 2 f(\epsilon_{\vec{p}})}{2 \epsilon_{\vec{p}} } \right] \nonumber \\
&=&  \rho \log \left[ A/ \Omega_0 \right] + \mathcal{O}(\Omega_0)
\end{eqnarray}
where $G_0(p)$ is the non-interacting Matsubara Green function
\begin{equation}
G_0(p) = \frac{1}{i p_0 - \epsilon_{\vec{p}}}.
\end{equation}
Had we been interested in the negative U Hubbard model, $\Gamma_s(L_a)$ and $\Gamma_s(L_b)$ are 
the most important quantities and give rise to the BCS instability in the trivial s-wave channel.  Note that 
$\Gamma(L_b)$ eliminates the dependence on the initial cutoff $\Omega_0$ as discussed in section VI.

For unconventional superconductivity that derives from repulsive interactions, we will need to consider 
higher order diagrams.  The next set of diagrams which are most important correspond to the set of 
diagrams in (L$_c$) of Fig. \ref{ladder}.  Of these the second-order vertex shown in (2a) corresponds to
\begin{eqnarray}
\Gamma_s(2a) &=& \sum_p G_0(p)G_0(k+q+p) \nonumber \\
&=&  \chi(\vec{k} + \vec{q} ) + \mathcal{O}( \Omega_0)
\end{eqnarray}
Adding the two second-order terms, we find the expression in Eq. \ref{Gamma2s}

The most divergent third order term is given by diagram (L$_d$) in Fig. \ref{ladder} which is simply 
\begin{equation}
\Gamma_s(L_d) =  \left[ \sum_p G_0(p) G_0(-p) \right]^2 =  \rho^2 \log^2 \left[ A/\Omega_0 \right]
\end{equation}
The next most important contribution comes from the diagrams in (L$_e$) and (L$_f$) with the thick solid line 
treated to second order (i.e. using diagram (2a)).  These produce 
\begin{eqnarray}
\Gamma_s^{(3)}(L_e) &=& \sum_{p p'} G_0(q+p+p')G_0(p') G_0(p)G_0(-p) \nonumber \\
&=& \left[ \sum_p  \chi(\vec{q} + \vec{p}) \left[G_0(p) G_0(-p)\right] + \mathcal{O}(\Omega_0) \right]\nonumber \\
&&+ \delta \Gamma_s(L_e) 
\end{eqnarray}
with
\begin{equation}
\delta \Gamma_s(L_e) = \sum_p \left[ \chi(\vec{q}+ \vec{p};ip_0)-\chi(\vec{q}+\vec{p},0) \right]G_0(p)G_0(-p)
\end{equation}
and the integrals in $\delta \Gamma_s(L_e)$ are carried out with $\Omega_0 = 0$, since this 
quantity is non-singular, and $\chi(\vec{p};ip_0)$ is the frequency-dependent susceptibility.  
In practice, this quantity will be evaluated numerically.  
Upon summing over the Matsubara frequencies and making use of the identity 
\begin{equation}
\label{measure}
\int_{\Omega_0} \frac{d^d p}{(2 \pi)^d} = \int \frac{d \hat{p}}{S_F} \frac{\bar{v}_F}{v_F(\hat{p})} \int_{\vert \epsilon \vert >  \Omega_0} d \epsilon \rho(\epsilon)
\end{equation}
where $\epsilon$ is the energy relative to the Fermi surface, it follows that 
\begin{equation}
\Gamma_s^{(3)}(L_e) =  \gamma^{(3)}(\hat{q}) \rho \log \left[ A/ \Omega_0 \right] + \delta \Gamma_s(L_e)
\end{equation}
The remaining third order terms are non-singular and are obtained from diagram (L$_c$) again, but this time, keeping only 3rd order contributions to the thick solid line, which are obtained from diagrams (3a)-(3g).  
While diagram (3a) is simply 
\begin{eqnarray}
\Gamma_s(3a) &=& \sum_{p p'} G_0(p) G_0(k+q+p)G_0(p') G_0(k+q+') \nonumber \\
&=& \chi^2(\vec{k}+\vec{q}) + \mathcal{O}(\Omega_0), 
\end{eqnarray}
diagrams (3b,3c) which are 
\begin{equation}
\Gamma_s(3b)=\Gamma_s(3c) = \sum_{p p'} G_0(p) G_0(q+p'-p) G_0(p') G_0(k+q+p')
\end{equation}
are not expressible in simple closed form due to the internal frequency 
integration (but it is possible to obtain these quantities to arbitrary precision 
in a numerical computation).  
Diagram (3d) contributes 
\begin{equation}
\Gamma_s(3d) =  \chi^2(\vec{k} - \vec{q})
\end{equation}
which is also non-singular, and lastly, the diagram of the form given in (3e) contributes 
\begin{equation}
\Gamma_s(3e) = - \sum_{p p'} G_0(p) G_0(p+k-q)G_0(p') G_0(p'-k-p)
\end{equation}

Next, we consider the fourth-order terms.  Again, the most singular contribution 
comes from the ladder shown in (L$_g$) of Fig. \ref{ladder}: 
\begin{equation}
\Gamma(L_g) = \rho^3 \log^3 \left[A/ \Omega_0 \right] + \mathcal{O}(\Omega_0)
\end{equation} 
The next most important contributions to this order comes from the 
diagrams in (L$_h$,L$_i$) with the thick solid line treated again only to second order, 
namely from diagram (2a).  Diagrams (L$_h$)  and (L$_i$) have the following contributions 
to leading logarithmic order:
\begin{eqnarray}
\Gamma_s(L_h) =  \rho^2 \log^2 \left[ A/ \Omega_0 \right] \gamma^{(3)}(\hat{k}) \nonumber \\
\Gamma_s(L_i) =  \rho^2 \log^2 \left[ A/ \Omega_0 \right] \gamma^{(3)}(\hat{q}) 
\end{eqnarray}
where again, we have neglected the frequency dependence of the susceptibility (which does not produce a singular contribution), and have made use of the identity in Eq. \ref{measure}.  Note that 
$\Gamma_s(L_h) $ depends only on the incoming momenta, whereas $\Gamma_s(L_i)$ depends only on 
the outgoing momenta.  In this way, it is easy to see that diagram (L$_j$) has no momentum dependence whatsoever:
\begin{equation}
\Gamma_s(L_j) =  \rho^2 \log^2 \left[A/\Omega_0 \right] \gamma_1^{(4)}
\end{equation}
The next leading terms are obtained again from the diagrams (L$_e$, L$_f$), except now, we require that the thick 
black lines in these diagrams take the non-singular third-order vertex which is given by $\tilde \Gamma_s^{(3)}$ described above:
\begin{equation}
\Gamma^{(4)}_s({\rm v}) = \rho \log \left[ A/ \Omega_0 \right] \gamma_2^{(4)}(\hat{q})  + \mathcal{O}(\Omega_0)
\end{equation}
and a similar expression holds for $\Gamma^{(4)}_s(L_f) $.  The final term which is singular at fourth order is given by the diagram (L$_k$), where the thick solid line is approximated by the second-order contribution in (2a).  This contributes 
\begin{eqnarray}
\Gamma_s(L_k) &=&  \sum_p \chi(\vec{k} + \vec{p}) G_0(p) G_0(-p) \chi(\vec{p} + \vec{q}) \nonumber \\
&=& U^4 \rho \log\left[ A/ \Omega_0 \right] \gamma^{(4)}_s(\vec{k}, \vec{q}) +\mathcal{O}(\Omega_0)
\end{eqnarray}
The remainder of terms that occur to fourth order do not make a singular contribution to the vertex.  
We shall not discuss them here.  

Now, we consider the triplet channel which has considerably fewer diagrams.  The lowest order contribution comes from the second-order diagram in (2b):
\begin{equation}
\Gamma_t(2b) = - \chi(\vec{k}- \vec{q})+\mathcal{O}(\Omega_0)
\end{equation}
The third order correction to the vertex comes from diagram (3g) which is 
\begin{eqnarray}
\Gamma_t(3g) &=& -   \sum_{p p'} G_0(p) G_0(k-q+p) G_0(p') G_0(k+p+p') \nonumber \\
&=& - \sum_p G_0(p) G_0(k-q+p) \chi(k+p)+\mathcal{O}(\Omega_0)
\end{eqnarray} 
Lastly, the fourth-order term comes from diagram (L$_n$), treating the thick solid line to $\mathcal{O}(U^2)$, ie. using diagram (2b).  This produces 
\begin{equation}
\Gamma_t(L_n) =  \rho \log\left[A/ \Omega_0 \right] \gamma_t^{(4)}(\vec{k}, \vec{q})+\mathcal{O}(\Omega_0)
\end{equation}
This contribution, although higher order, has an important conceptual importance: it has the correct logarithmic dependence on $\Omega_0$ to remove the dependence of the final expression for $T_c$ 
on the initial cutoff.

\bibliography{smallU}

 \end{document}